\documentclass[12pt]{article}
\input{amssymb.sty}
\input epsf.def

\def\thefootnote{\fnsymbol{footnote}}
\catcode`\@=11
\def\numberbysection{\@addtoreset{equation}{section}
         \def\theequation{\thesection.\arabic{equation}}}
\numberbysection

\def\beq{\begin{equation}}
\def\eeq{\end{equation}}
\def\bea{\begin{eqnarray}}
\def\eea{\end{eqnarray}}
\def\bdis{\begin{displaymath}}
\def\edis{\end{displaymath}}
\def\cqfd{\hskip 1truemm \vrule height2mm depth0mm width2mm}

\def\z{\zeta}

\def\Q{{\Bbb Q}}
\def\Z{{\Bbb Z}}
\def\N{{\Bbb N}}
\def\H{{\cal H}}
\def\R{{\cal R}}
\def\F{{\mathbb F}}
\def\G{\cal{G}}

\renewcommand{\rho}{\varrho}
\begin{document}

\pagenumbering{alph}
\setcounter{page}{0}

\rightline{UCL--IPT--00--06}

\vskip 4cm
{\LARGE \centerline{On Discrete Symmetries in su(2) and su(3)}
\centerline{Affine Theories and Related Graphs}}

\vskip 3cm

\centerline{\large S. Li\'enart\footnote{Chercheur FRIA}, P. 
Ruelle\footnote{Chercheur
Qualifi\'e FNRS} and O. Verhoeven\footnote{Collaborateur scientifique FNRS}}

\vskip 2truecm
\centerline{Institut de Physique Th\'eorique}
\centerline{Universit\'e Catholique de Louvain}
\centerline{B--1348 \hskip 0.5truecm Louvain-La-Neuve, Belgium}

\vskip 3truecm
\begin{abstract}
\noindent
We classify the possible finite symmetries of conformal field theories with
an affine Lie algebra $\widehat{su}(2)$ and $\widehat{su}(3)$, and discuss the results
from the perspective of the graphs associated with the modular invariants. The
highlights of the analysis are first, that the symmetries we found in either case are
matched by the graph data in a perfect way in the case of $su(2)$, but in a looser
way for $su(3)$, and second, that some of the graphs lead naturally to projective
representations, both in $su(2)$ and in $su(3)$. 
\end{abstract}

\renewcommand{\thefootnote}{\arabic{footnote}}
\setcounter{footnote}{0}
\newpage
\pagenumbering{arabic}
\baselineskip=16pt


\section{Introduction}

Since 1984 \cite{bpz}, two--dimensional conformal field theory has remained a
fascinating subject, at the frontier of mathematics and physics. In two--dimensional 
physics, conformal theories proved to be a powerful tool in the study of critical and
off--critical models, and in the dynamics of string theory through the description of
the world--sheet. More recently, they have prompted an overwhelming activity in
mathematics, in as different subject as the theory of singularities, knot theory,
combinatorics, algebraic geometry, von Neumann algebras, ...

Strangely, many of the developments have been triggered by the problem of modular
invariance, most notably by the remarkable and not yet fully understood ADE
classification of $su(2)$ affine theories. Affine theories are conformal field theories
with an affine Lie algebra as symmetry algebra, prime examples being the
Wess--Zumino--Witten models (see \cite{g,w} for recent reviews). Despite much progress
and many partial results, only the simpler cases have been fully classified: those
based on the $\widehat{su}(2)$ \cite{ciz,kato} and $\widehat{su}(3)$ \cite{gannon}. 

The modular invariants are torus partition functions with periodic boundary conditions
in both directions. A better understanding of these models can be gained by allowing
other types of monodromies, a systematic source of which being twists effected by the
elements of an internal symmetry group. As shown by Zuber \cite{zuber}, this can 
be done from the knowledge of the critical modular invariants, by examining
the modular covariance of the sought twisted partition functions. Illustrations of the
method were given in the $(A,A)$ series of Virasoro minimal models, but
the analysis was subsequently completed and extended to all minimals models
\cite{rv}, with the somewhat expected results that the symmetry group ---discrete--- of
a model $(A,{\cal G})$ is simply the automorphism group of the Dynkin diagram of
$\cal G$. This result feeds the general feeling that the graphs that have been
associated with  modular invariants are key elements in the description of the
corresponding conformal theory, a bit like a Dynkin diagram commands a Lie algebra.

It is our purpose to pursue this analysis for affine models, based on the affine
algebras $\widehat{su}(2)$ and $\widehat{su}(3)$. For those, we determine the maximal
discrete symmetry group and compute the partition functions in all twisted sectors.
There are several motivations for doing this.

First of all, no affine model yet has been realized as the scaling limit of a
critical lattice model, unlike the unitary minimal Virasoro models. Thus the knowledge
of the symmetries can be a useful guide in the search of such models. In addition, the
modular invariant partition function contains information on the periodic sector only,
and so cannot distinguish between theories with the same partition function but
otherwise distinct. Symmetry arguments can do this, or can at least suggest this
possibility, as in the $\widehat{su}(3)$ models at level $k=3$ and 6, see below.
Moreover, the presence of a symmetry implies selection rules in the operator product
algebra, both in the bulk and on boundaries, resulting in non--trivial constraints on
operator product coefficients \cite{r}. And last, the relevance of the graphs may be
further probed, by comparing the symmetries found in the affine theory and those of the
associated graph(s). From this point of view, we will see that, even though the results
are qualitatively different for the $\widehat{su}(2)$ and $\widehat{su}(3)$ models, the
graphs are definitely relevant also for symmetries.

The plan of this article is as follows. In Section 2, we first recall the general
setting for discrete symmetries and twisted boundary conditions. We explain the
strategy we use in order to determine the maximal symmetry compatible with a given
modular invariant, and also the catches of such an analysis. After a few words about
orbifolds, we recall the salient features of the association of graphs with modular
invariants. 

In the following section, we summarize our results, in the form of lists of groups and
of partition functions, for all $su(2)$ and $su(3)$ modular invariant theories. We
then undertake a general comparison of these results with the graph data. A study of
particular cases, which we found the most intructive or representative ---all taken from
$su(3)$ models---, finishes the third section. 

Section 4 is devoted to the proofs. Many of the above results can be proved using the
techniques of \cite{rv}. So we will content ourselves with giving the explicit proof in
two representative series of modular invariants (the type II $D$--series of automorphism
invariants of $su(2)$, and the type I $D$--series of $su(3)$). 

A general summary concluding this work is presented in Section 5. 


\section{Symmetries and Graphs}

\subsection{Frustrated partition functions}

A particularly interesting geometry to study two--dimensional conformal field theory
is the toroidal geometry. It is well--known that two different tori are conformally
equivalent if they are related by a modular transformation.
Explicitely, if we denote by $ \tau $ (Im  $ \;  \tau > 0$) the 
standard modulus of the torus, the action of the modular group is given by
\beq
\Big\{\tau \rightarrow {a\tau + b \over c\tau + d} \;:\;
a,b,c,d \in \Z \; {\rm and} \; ad-bc=1\Big\}
\eeq
This group, isomorphic to PSL$_2(\Z)$, is generated by the two transformations
$T\;:\; \tau \rightarrow \tau+1$ and $S\;:\; \tau \rightarrow -{1  \over \tau}$.

A conformally invariant theory on the torus (i.e., with doubly periodic boundary
conditions) is thus described by a modular invariant partition function (MIPF). Here,
we consider theories based on untwisted affine Lie algebras \cite{kac}, which implies
that the MIPFs are sesquilinear forms in the characters, 
\beq
Z(\tau) = \sum_{p,p'} \; [\chi_{p}(\tau)]^* \, M_{p,p'} \, [\chi_{p'}(\tau)], 
\label{mipf}
\eeq
for non--negative integral coefficients $M_{p,p'}$. The sum over the character labels
$p,p'$ is finite.

The classification of all modular invariant partition functions has been completed for 
the algebras $\widehat{su}(2)_k$ \cite{ciz,kato} and $\widehat{su}(3)_k$ \cite{gannon}.
The classification for the other algebras is so far incomplete, which is the reason why
we focus in this article on $su(2)$ and $su(3)$.

While MIPFs are associated with periodic boundary conditions, there may be other
choices of boundary conditions, corresponding to different monodromy properties of the
fields along non--contractible loops. In particular, twisted, non--periodic boundary
conditions are possible in a theory which has a symmetry group. The group, $G$ say,
acts on the various fields $\phi \rightarrow {}^g\,\phi$ and leaves the (periodic,
bulk) Hamiltonian invariant. The twisted boundary conditions induced by the action of
$G$ can be written, in the case of the torus, as
\beq
\phi(z+1)={}^g\phi(z), \qquad \phi(z+\tau)={}^{g'}\!\phi(z), \qquad 
\hbox{ with $g,g' \in G$}.
\eeq
A pair $(g,g')$ defines a specific sector of the theory corresponding to the
given twisted boundary conditions. The consistency of the boundary conditions with the
translation group of the torus requires that $g$ and $g'$ commute. 

Each sector of boundary conditions has its own partition function $Z_{g,g'}$ (called
frustrated or twisted in the case of non--periodic b.c.), given, in the Hamiltonian
formalism, by \cite{cardy2,iz}
\beq
Z_{g,g'}(\tau) = {\rm Tr}_{\H_g} \, \Big(q^{L_0-c/24} \, \overline
q^{\overline L_0-c/24}g'\Big), \qquad q=e^{2i\pi \tau}.
\eeq
This formula gives a different status to $g$ and $g'$. $\H_g$ is the Hilbert space of
states that live on the fixed time slices (chosen to be lines of constant Im$\,z$ in the
complex plane), and is subjected to the $g$ boundary condition, while the boundary
condition in the time direction is effected by the insertion of $g'$. Thus $g$
specifies the $g$--sector $\H_g$ of states/fields of the theory, which carries an
action of the centralizer $C(g)$ of $g$ (to which $g'$ belongs). This apparent
asymmetry between $g$ and $g'$ cannot exist since there is no preferred direction on
the torus. The restoration of the symmetry, based on the modular covariance, is 
the basis for the determination of the symmetry group of a particular theory. 

We have mentioned that $g'$ must be in the centralizer $C(g)$ of $g$, which thus
appears as the symmetry group that remains unbroken in the $g$--sector (so the periodic
sector ($g=e$) has the maximal symmetry, equal to $G$ itself). It is also easy to
see that the partition function $Z_{g,g'}$ depends only on the conjugacy class of $g'$
within $C(g)$, and on that of $g$ (within $G$). Eliminating all redundancies, we obtain
a (in general non--square) list of frustrated partition functions $Z_{g,g'}$.

We finish this introductory section by recalling the compatibility conditions that
follow from putting together the modular transformations and the existence of a
symmetry group. 

The presence of the affine symmetry implies that every space $\H_g$ is made up of
representations of a pair $\widehat{\G}_k \times \widehat{\G}_k$, with $\widehat{\G}_k$
the level $k$ affine Lie algebra based on $\G$ ($=su(2)$ or $su(3)$ here)
\beq
\H_g = \bigoplus_{p,p' \in P_{++}^{(n)}} \; M_{p,p'}^{(g)} \; (\R_p \otimes \R_{p'}),
\eeq
where the numbers $M_{p,p'}^{(g)}$ are multiplicities, {\it i.e.} non--negative
integers. The shifted weights $p,p'$ specify the inequivalent integrable representations
of $\widehat{\G}_k$, and the height $n$ is defined by $n=k+h^\vee$, with $h^\vee$ the
dual Coxeter number of $\G$.

Assume now that the theory has a symmetry group $G$. It means that the generators of
the affine Lie algebra (and of the Virasoro algebra) are $G$--singlets. This
in turn implies that the action of $C(g)$ is block--diagonal on inequivalent
representations occurring in $\H_g$, so that, for each pair $p,p'$, the
$M_{p,p'}^{(g)}$ equivalent affine representations carry a representation of $C(g)$.
Using the definition of the affine characters $\chi_{p}(q) = {\rm
Tr}_{\R_p}\, q^{(L_0-c/24)}$, one obtains\footnote{Provided a proper modification of
the stress--energy tensor is made, the same can be done with unrestricted affine
characters \cite{bppz}.}
\beq
Z_{g,g'}(\tau) = \sum_{p,p' \in P_{++}^{(n)}} \;
M_{p,p'}^{(g)} (g') \; \chi^*_p(q) \,\chi^{}_{p'}(q),
\label{tpf}
\eeq
where $M_{p,p'}^{(g)} (g')$ is the character of a unitary representation of $C(g)$
of degree $M_{p,p'}^{(g)} \equiv M_{p,p'}^{(g)} (e)$. As an additional requirement, we
will impose that the identity (label $p=\varrho$) appears in the periodic sector
$\H_e$ only and is invariant under $G$, so that $M_{\varrho,\varrho}^{(g)} (g') =
\delta_{g,e}$.

For a fixed $g'$, say of order $N$, this representation can be fully diagonalized,
which then reveals the charges, defined modulo $N$, of the individual representations 
\beq
M_{p,p'}^{(g)} (g') = \sum_{k=1}^{M_{p,p'}^{(g)}}  \;\z_N^{Q(g;p,p';g';k)},
\eeq
with $\z_N = e^{2i\pi/N}$. In most cases, the multiplicities $M_{p,p'}^{(g)}$ are equal
to 1, so that only one--dimensional representations are involved, from which 
charges can be read off in a straightforward way.

The frustrated partition functions (\ref{tpf}) are subjected to strong constraints
since they must be compatible with the modular transformations. These leave the torus
invariant but mix the two periods, and hence the boundary conditions. The result is the
following transformation formula 
\beq
Z_{g,g'}(\tau) = Z_{g,gg'}(\tau+1) = Z_{g'^{-1},g}({\textstyle -{1 \over
\tau}}) = Z_{g^ag'^c,g^bg'^d}({\textstyle {a\tau+b \over c\tau+d}}). 
\label{tmod}
\eeq
These are severe constraints, which allow to determine the possible symmetries (and
all frustrated partition functions) compatible with a given theory (a given MIPF).  

In practice, one starts from a supposedly known MIPF $Z_{e,e}$. From it one tries to
determine all $Z_{e,g'}$, whose form (in terms of affine characters) is fixed by
$Z_{e,e}$, and which themselves yield other functions $Z_{g,g'}$ by modular
transformations. This set of functions is finally extended in a maximal way so as to
obtain a table of functions $Z_{g,g'}$ which transform in the covariant way prescribed
by (\ref{tmod}). The resulting table corresponds to the maximal symmetry group $G$ that
is compatible with the theory specified by the initial MIPF. This procedure has been
used in \cite{rv} in the case of minimal Virasoro models.

It must be emphasized that we assume throughout that the symmetry group preserves the
affine algebra, which implies that all twisted sectors carry representations of the same
symmetry algebra. With this in mind, the non--existence of a symmetry group can mean
two things. Either the theory really has no symmetry, or it does have one but which
does not preserve the level of the chiral symmetry algebra from which we have chosen to
look at the theory, in which case a lower, subalgebra point of view must be taken.  


\subsection{The determination of the symmetry group}

The end of the previous section sketched the way the maximal symmetry group of a theory
can be found. Here, we make the procedure a bit more explicit, but more importantly, we
discuss the various issues that need to be clarified before the symmetry group can be
safely named.

The most obvious question is related to the fact that only pairs $(g,g')$ of commuting
elements can be used. In a sense, the non--abelian features of the group cannot all be
probed directly.

A good starting point is to classify the cyclic symmetries, and their possible
realizations ---there can be more than one---, according to the scheme described above.
Being abelian groups, there is no restriction on $g,g'$, and their full structure can be
exposed. Since most theories (in this paper) have an abelian cyclic symmetry, this will
be the complete story. In a few cases, distinct cyclic symmetries or several
realizations of a given cyclic symmetry will be found, indicating the existence of a
larger group. 

Two cyclic symmetries $\Z_N$ and $\Z_{N'}$ can be assembled into $\Z_{NN'}$ if an
action of $\Z_N$ can be consistently defined in the sectors labelled by the elements of
$\Z_{N'}$, and vice--versa, i.e. if the two kinds of charges are simultaneously
assignable. If not, the two cyclic factors are not commuting subgroups. In this way,
one can list the maximal abelian subgroups and look for non--abelian groups that
contain them (and only them). In case several groups qualify, one should be able
to pick the right one by looking at the various sectors $\H_g$ and the associated
data $M_{p,p'}^{(g)}(g')$, which are to be characters of the centralizer $C(g)$. 

This procedure presumably leads to a unique group $G$, as it has to match a number of
data: the conjugacy classes and their centralizers, the maximal abelian subgroups, and
the character tables of the centralizers (maybe not the whole of them). 

Subtleties or difficulties can however be encountered in the above analysis. 

\begin{enumerate}

\item
In looking for a cyclic symmetry, we usually proceed by identifying the way it
can be realized in the periodic sector. This yields the functions $Z_{e,g}$ which, by
modular transformations, allow to compute the complete table of partition functions
$Z_{g,g'}$ for $g,g' \in \Z_N$. When there is a unique function $Z_{e,g}$, this
procedure produces a unique table of functions $Z_{g,g'}$. This is however not enough
to guarantee that the group $\Z_N$ itself is unique, and indeed it sometimes extends to
a power $\Z_N^m$.

If there is a $\Z_N$ action in the periodic sector by one--dimensional representations
(i.e. all fields appear with a degeneracy equal to 1), then $Z_{e,g}$ gives the exact
way the generator of $\Z_N$ acts. So if there is a unique function $Z_{e,g}$, there is
also a unique $\Z_N$ action, and a unique $\Z_N$.

This is no longer the case if higher--dimensional representations are involved, since
$Z_{e,g}$ sees only their trace. Suppose that, among others, $\Z_N$ acts
in the periodic sector by an $M$--dimensional representation $R_M$ (a field has
multiplicity $M$; here, $M \leq 3$ for $su(2)$ and $su(3)$). That the functions
$Z_{e,g}$ are unique means that, restricted to that block, any other $\Z_N$ must act by
an $M$--dimensional (distinct) representation $R'_M$ that has the same character as
$R_M$.

So if there is, say, a $\Z_N \times \Z_N$ symmetry, all $\Z_N$ subgroups must act on
this $M$--by--$M$ block with the same character. As it turns out, this implies that $M$
must be large enough for this to happen (as we have seen above, $M$ cannot be 1), with
in addition restrictions on the character. For $N=2$, $M$ must be at least equal to 3,
and the character of the representations must be equal to $-1$. Written in 
diagonalized form, the representations of the three $\Z_2$ subgroups of $\Z_2 \times
\Z_2$ (generated by $g_1$ and $g_2$) read : $R(g_1) = {\rm diag}(1,-1,-1)$, $R(g_2)
= {\rm diag}(-1,1,-1)$ and $R(g_1g_2) = (-1,-1,1)$. Likewise, for $N=3$, the number $M$
must be bigger than 8, and the representations must also have a trace equal to
$-1$\footnote{For general $N$, a solution exists with $M=N^2-1$ and again a trace equal
to $-1$. We suspect that this value of $M$ is minimal.}.

As mentioned above, $M$ is at most 3 in the models considered here, so that $\Z_2$ is
the only symmetry that can potentially be extended, requiring in addition a trace equal
to $-1$. But the only model which has a $\Z_2$ symmetry and a multiplicity equal to 3 is
$\widehat{su}(3)_3$ (height 6), for the partition function $D^{}_6 = D^*_6$, where a 
cyclic symmetry $\Z_2$ indeed leads to a $\Z_2 \times \Z_2$ group, the two factors being
conjugate within a bigger group, namely $A_4$. The action is precisely given by the
three matrices given above. The details for the determination of the functions
$Z_{g_1^kg_2^\ell, g_1^{k'}g_2^{\ell '}}$ are given in Section 3.3.1.

\item
The last step in the procedure described above may not work. Indeed the list of
all maximal abelian subgroups poses no problem, but there is no guarantee that a single
group can accomodate them all. In other words, there may not be a unique maximal
symmetry group, but several ones (possibly isomorphic). That this occurs can be taken
as an indication that different models exist, which all share the same modular invariant
partition function. 

The only case where we have seen this situation occur is again in $\widehat{su}(3)_3$,
for the partition function $D^{}_6 = D^*_6$. There we found two isomorphic
maximal symmetry groups $G = G' = A_4$, the alternating group on four letters. The two
groups
$A_4$ have $\Z_3$ and $\Z_2 \times \Z_2$ as maximal abelian subgroups, but carry a
different realization of the $\Z_3$ factor. The existence of these two symmetry groups
is paralleled by the existence of two isospectral graphs, naturally associated to the
modular invariant partition function, and reproducing correctly the characters found in
the periodic sector $\H_e$ (see Section 3.3.1 for more details). 

These elements give some support to the existence of two different models with
the same torus partition function $D^{}_6 = D^*_6$.

\item
The last issue we want to mention is the possibility that the symmetry group acts on
the various sectors by projective representations. This possibility can be easily
accounted for if one bears in mind that a projective action modifies the modular
transformations of the partition functions.  

Let us suppose that $G$, in fact $C(g)$, acts on the sector $\H_g$ by a projective
representation $R_g$, characterized by a 2--cocycle $\omega_g$:
\beq
R_g (h) R_g (h') = \omega_g(h,h') R_g(h\,h'), \quad \forall h,h' \in C(g).
\eeq

For such representations, the partition functions $Z_{g,g'}$ transform differently from
(\ref{tmod}), since the composition of group elements is involved. It is not
difficult to see that the transformation rules are now
\beq
Z_{g,g'}(\tau) = \omega_g (g,g') \; Z_{g,gg'}(\tau + 1) = 
\omega_{g'}^{-1}(g,g^{-1}) \; Z_{g',g^{-1}} ({\textstyle -{1 \over \tau}}).
\label{pmod}
\eeq

Let us remark that one should in general expect the various cocycles $\omega_g$ to be
cohomologically trivial. In the sector ${\cal H}_g$, the action of $C(g)$  is block
diagonal with respect to the inequivalent affine representations. So, unless all
occurring representations have multiplicity bigger than 1, some of the blocks will
correspond to one--dimensional $\omega_g$--projective representations. This implies
that $\omega_g$ is a coboundary. 

Although we were not so much interested in projective representations, they were somehow
forced on us in a number of cases, where an expected symmetry, present in the graphs,
was not seen to be realized in the field theory, but could however be realized
projectively. This situation occurs for a $\Z_2$ group in all diagonal $su(2)$ models
for an odd height (the $A_{n-1}$, $n$ odd, theories), and the $E^{(*)}_8$ and
$E_{24}^{}$ models of $su(3)$. In the $E_8^{(*)}$ models of $su(3)$, it combines with a
$\Z_3$ symmetry to form a projective $\Z_6$. To our knowledge, this is the first 
instance of such a situation. 

\end{enumerate}


\subsection{Orbifolds}

The orbifold construction is well--known (see for instance \cite{dms}), and roughly
speaking, corresponds to quotienting a theory $\cal T$ by its symmetry group $G$ (or
subgroup). The resulting, orbifold theory ${\cal T}/G$ may or may not be different from
$\cal T$ itself. 

In some cases, most notably when the group $G$ is abelian, the orbifold theory share
the same symmetry as the original one, which can be recovered by orbifolding the
orbifold theory. So the two theories, considered with all their twisted sectors, are
essentially equivalent. 

When $G$ is non--abelian, the symmetry of the orbifold theory is smaller, and equal to
the abelianization of $G$, namely $G/G'$, where $G'$ is the commutator subgroup
\cite{ginsp}. The torus modular invariant partition function of the orbifold theory,
\beq
Z_{e,e}^{\rm orb} = {1 \over |G|} \sum_{g \in G} \sum_{g' \in C(g)} Z_{g,g'} = 
{1 \over |G|} \sum_{[g]} |[g]| \, \sum_{g' \in C(g)} Z_{g,g'},
\label{orb}
\eeq
collects, from all sectors of the original theory, the fields which are singlets under
$G$ ($|[g]|$ is the cardinal of the class of $g$, and the first summation is over the
classes of $G$). 

The formula (\ref{orb}) is not the only way to obtain a modular invariant
partition function. Distinct orbits under the modular group can be given different
coefficients and the result is still modular invariant. That the
invariant is a modular invariant partition function puts further restrictions which have
been examined in \cite{vafa}: the different possibilities are classified by the second
cohomology group $H^2(G,U(1))$ and lead to what is known as discrete torsion.

Take a 2--cocycle $\omega$ on $G$, and define 
\beq
\epsilon_g(g') = \omega(g,g') \, \omega(g',g)^{-1}.
\eeq
For each $g$, the function $\epsilon_g(\cdot)$ is a (true) representation of $C(g)$,
which can then be used to define the orbifold modular invariant partition function
\beq
Z_{e,e}^{\rm orb} = {1 \over |G|} \sum_{g \in G} \sum_{g' \in C(g)}
\epsilon_g(g') \, Z_{g,g'}.
\label{torb}
\eeq
Clearly, this new function corresponds to a projection, in each sector $\H_g$, onto the
fields which transform according to $\epsilon^*_g$.

In this paper, our emphasis is not on the orbifold construction, except that
we use it as a cross--check on the frustrated partition functions we find. At
the same time, we will compare what the orbifold procedure gives at the level of the
graphs, confirming that the graph associated to the orbifold theory is the orbifold
graph. 

The only model where there is some room for discrete torsion is, once more, the
$D^{(*)}_6$ models of $su(3)$, since the symmetry group of all other models has a
trivial cohomology $H^2$. Details about these models are given in Section 3.3.1, where
the discrete torsion yields nothing new compared to the ordinary orbifold construction. 


\subsection{Graphs}

Graphs lie at the heart of the structure of the $su(2)$ theories. Let us recall that
each $su(2)$ modular invariant $Z$ can be uniquely associated with a graph $\Gamma$, in
this case an ADE Dynkin diagram, such that the diagonal terms of $Z$ can be recovered
from the graph spectral data \cite{ciz}. Although the graph does not specify, in a
direct way, the whole modular invariant, it does give it implicitely through the modular
invariance itself (no distinct modular invariants are known that share the same
diagonal terms). It remains a remarkable observation that the full set of consistent
$su(2)$ theories is simply the ADE list. The same observation applies to the Virasoro
minimal models, for which the graphs are in fact pairs of graphs, closely related
to an A Dynkin diagram and an ADE Dynkin diagram (with a restriction on the number of
nodes). 

The generalization of these ideas to other models has been considered in \cite{dfz,df}.
The starting point is always a MIPF, for which one tries to put the
diagonal\footnote{Other couplings than the diagonal ones may be relevant as well
\cite{fs2}.} terms in correspondence with a graph or a collection of graphs. As shown
the most clearly in \cite{bppz}, the connection gets even more significance if one
phrases it in terms of an $\N$--representation of the fusion algebra. We briefly recall
the correspondence (a more complete account of the many appearances of graphs in
various contexts can be found in \cite{z2}).

In a theory with identical chiral and antichiral algebras, let $i$ be a label for the
chiral primary fields, with the label 0 for the identity. The primary fields satisfy
a fusion algebra with positive integer structure constants $N_{ij}^k = N_{ji}^k$. The
matrices $(N_i)_j^k = N_{ij}^k$ themselves satisfy the fusion algebra, and are all
diagonalized by the modular $S$ matrix. As a consequence, the eigenvalues of $N_i$
are given by the set $\{S_{ij}/S_{0j}\}_j$. 

Let us now assume that $Z$ is the modular invariant partition function of that theory
on the torus. If we write the diagonal terms of $Z$ as 
\beq
Z = \sum_{i \in {\cal E}} |\chi_i|^2 + \hbox{non--diagonal},
\eeq
the set $\cal E$ contains all labels $i$ (with possible multiplicity) such that the
periodic sector contains the non--chiral scalar field $(i,i)$.

Given the set $\cal E$, one looks for a $\N$--representation $n_i$ of the
fusion algebra, of dimension $|{\cal E}|$, such that the eigenvalues of $n_i$ are
determined from $\cal E$:
\beq
\hbox{spec}\,n_i = \{{S_{ij} \over S_{0j}} \;:\; j \in {\cal E}\}.
\eeq

The regular representation of the fusion algebra, $n_i=N_i$, is always a solution,
that corresponds to $\cal E$ equal to the whole set of available labels, and thus to the
diagonal modular invariant. 

All matrices $n_i$ have positive integer entries, and so can be viewed as adjacency
matrices of graphs (having $|{\cal E}|$ nodes). Since the fusion algebra has usually a
restricted number of generators, the diagonal structure of $Z$ can eventually be
associated with a restricted number of fundamental graphs. For $su(2)$, one generator,
hence one graph, is sufficient. For $su(3)$, one needs two generators, and so two
(oriented) graphs, but the conjugation implies that the two adjacency matrices are the
transpose of each other (the two graphs differ by their orientation). Therefore, both in
$su(2)$ and $su(3)$, the information can be condensed in a single graph. We will denote
its adjacency matrix by $n_f$.

Remarkably, all $\N$--representations are known in the case of the $su(2)$ fusion
algebras \cite{dfz}. At height $n=k+2$, the only (irreducible) solutions correspond to
taking for $n_f$, the generator of the fusion ring, the adjacency matrix of all
ADE Lie algebras with Coxeter number equal to $n-1$ or the adjacency matrix of the
tadpole graph T with $n-1$ nodes. Only the ADE diagrams correspond to sets $\cal E$
which are realized in modular invariants. All representations based on T, and direct
sums of ADET diagrams are spurious in the sense that their set $\cal E$ does not
correspond to a modular invariant. Those solutions are thus discarded. 

For $su(3)$, the authors of \cite{dfz} propose a tentative list of graphs (see also
\cite{bppz}). For each modular invariant, they found at least one graph with the right
properties, but the novelty, as compared to $su(2)$, is that in a few cases, several
isospectral graphs exist. It is presently not known whether this list is
exhaustive.

In general, a representation $n_i$ may have a symmetry, in the sense that permutations
$\sigma$ of $\cal E$ leave the representation invariant
\beq
(n_i)_{\sigma(j)}^{\sigma(k)} = (n_i)_j^k, \qquad \forall i.
\eeq
When the representation is given in terms of a single matrix, or one graph, its
symmetry group is just the automorphism group of that single graph. In the regular
representation, the symmetries of the (fusion) graphs are given by simple currents.

One of our motivations for this work is precisely to see if the relevance
of the graphs can be further probed by taking the point of view of symmetries. As
far as graphs are concerned, our purpose will be to see if the symmetry data we find in
the conformal theories coincide with those we  get from the graphs. That question has
been answered positively\footnote{But for one class of models: the non--unitary minimal
models based on $(A_{p-1},A_{q-1})$ for $p$ and $q$ both odd. Strangely, the expected
$\Z_2$ symmetry does not even seem to be realized projectively.} in the case of the
minimal models, both on the torus \cite{rv} and on the cylinder \cite{r} where the role
played by the graphs in the way boundary conditions behave under the symmetry was
investigated. We will see here that it is still largely true in the $su(2)$ and
$su(3)$ models, though we found a few cases where the CFT and the graph picture do not
quite match. The highlights of this comparison are collected in Section 3.2.

We should close this section by stressing that the connection with graphs goes beyond
the diagonal terms of a MIPF, in at least five ways. First the graphs themselves can be
used to construct critical lattice models whose continuum limits yield conformal
theories which are filiated (via cosets) to the theory that was defining the graph
in the first place \cite{abf,pas,dfz,df,wns,o'bp,bp}. (No lattice models however are
known that lead in their continuum limit to the affine models themselves.) Second, in
the cases of type I (block--diagonal) invariants, a subset of nodes can be found within
the graph from which the block structure, hence the full invariant, can be recovered
\cite{df}. Third, the graphs leave a trace in some OPE structure constants \cite{pz}.
Fourth, the non--diagonal pieces of a modular invariant have also received an
interpretation in the context of von Neumann factors \cite{bek}. And
fifth, the graphs and the corresponding fusion representation $n_i$ play a central role
on the cylinder, where they classify the possible boundary conditions
\cite{cardy3,pss,fs1,bppz}. 


\section{Results}

This section presents the results of our investigation regarding the symmetries of
the $su(2)$ and $su(3)$ affine models. The first subsection makes a list of the symmetry
groups for the various theories, as well as their frustrated partition functions, as
obtained from the algebraic program described above. In a second part, these results
are viewed from the general graph theoretic perspective: without going into the details
of all the models, we examine to what extent the symmetry of the theories match that of
the graph, and draw general conclusions, mostly based on observations, on what the
graphs are capable to say about the conformal theories. The last part is devoted to
giving some details about the few cases we found the most instructive. Proofs are
relegated to the subsequent section.

We stress the fact that the groups $G$ listed below represent the maximal symmetries
which can have a non--projective realization in the corresponding field theories. We
have not determined the maximal symmetries with projective realizations. The only
projective group actions we have looked at are those connected to automorphisms of the
associated graphs $\Gamma$. So for a given theory and its graph, we denote by $G_{\rm
proj}$ the maximal subgroup of Aut$\, \Gamma$ that can be realized projectively. 

\subsection{Lists}

{\it (a) \underline{su(2) models}}

\bigskip \noindent
The results we found regarding the symmetries of the ${su(2)}$ are as follows.


\renewcommand{\arraystretch}{1.8}
\begin{table}[tpb]
\hspace{-5mm}
\begin{tabular}{|c|l|}
\hline
 & $\Z_2$--frustrated $su(2)$ partition functions ($g^2 = e)$ \\
\hline \hline
$\matrix{A_{n-1} \cr \mbox{{\footnotesize $n \geq 3$}}}$ & 
\begin{tabular}[t]{ll}
$Z_{e,e}= \,{\displaystyle  {\sum_{p=1}^{n-1}\;|\chi_{p}|^2}}$ &
$Z_{e,g}= \,{\displaystyle {\sum_{p=1}^{n-1}\, (-1)^{p+1} \; 
|\chi_{p}|^2}}$\\
$Z_{g,e} = \,{\displaystyle 
{\sum_{p=1}^{n-1}\;\chi^*_{p}\,\chi^{}_{n-p}}}$ &
$Z_{g,g} = \,{\displaystyle {\sum_{p=1}^{n-1}\,
(-1)^{p+n/2}\;\chi^*_{p}\,\chi^{}_{n-p}}}$\\
(Projective realization for $n$ odd) & \\
\end{tabular}\\
\hline
$\matrix{D_{{n \over 2}+1} \cr \mbox{{\footnotesize $n=4m+2$}}}$ & 
\begin{tabular}[t]{ll}
$Z_{e,e}= \,{\displaystyle \sum_{p=1,\, {\rm
odd}}^{2m-1}\, |\chi_{p}+\chi^{}_{n-p}|^2 + 2\, |\chi^{}_{2m+1}|^2 }$ &
\hspace{3mm} $Z_{e,g}= \,{\displaystyle \sum_{p=1,\, {\rm odd}}^{2m-1}
\,|\chi_{p}-\chi^{}_{n-p}|^2}$ \\
$Z_{g,e}= \,{\displaystyle \sum_{p=2,\, {\rm even}}^{2m}\,
|\chi_{p}+\chi^{}_{n-p}|^2}$ &
\hspace{3mm} $Z_{g,g}= \,{\displaystyle \sum_{p=2,\, {\rm even}}^{2m}\, 
|\chi_{p}-\chi^{}_{n-p}|^2}$ \\
\end{tabular}\\
\hline
$\matrix{D_{{n \over 2}+1} \cr \mbox{{\footnotesize $n=4m+4$}}}$ & 
\begin{tabular}[t]{ll}
$Z_{e,e}= \,{\displaystyle \sum_{p=1,\, {\rm odd}}^{n-1}\, |\chi_{p}|^2 +
\sum_{p=2,\, {\rm even}}^{n-2}\, \chi_{p}^*\,\chi^{}_{n-p}}$ &
\hspace{3mm} $Z_{e,g}=\,{\displaystyle \sum_{p=1,\, {\rm odd}}^{n-1}\,
|\chi_{p}|^2 - \sum_{p=2,\, {\rm even}}^{n-2}\, \chi_{p}^*\,\chi^{}_{n-p}}$ \\
$Z_{g,e}= \,{\displaystyle \sum_{p=2,\, {\rm even}}^{n-2}\, 
|\chi_{p}|^2 + \sum_{p=1,\, {\rm odd}}^{n-1}\, \chi_{p}^*\,\chi^{}_{n-p}}$ &
\hspace{3mm} $Z_{g,g}= \,{\displaystyle \sum_{p=2,\, {\rm even}}^{n-2}\, 
|\chi_{p}|^2 - \sum_{p=1,\, {\rm odd}}^{n-1}\, \chi_{p}^*\,\chi^{}_{n-p}}$ \\
\end{tabular}\\
\hline
$\matrix{E_6 \cr \mbox{{\footnotesize $n=12$}}}$ & $Z_{e,e}=
\,{\displaystyle \; |\chi_{1}+\chi_{7}|^2 +  |\chi_{4}+\chi_{8}|^2 +
|\chi_{5}+\chi_{11}|^2}$ \\
& \hspace{0.5cm} $Z_{e,g}= \,{\displaystyle  \; |\chi_{1}+\chi_{7}|^2 - 
|\chi_{4}+\chi_{8}|^2 +
|\chi_{5}+\chi_{11}|^2}$ \\
& $Z_{g,e}= \,{\displaystyle  \; |\chi_{4}+\chi_{8}|^2 + 
\Big\{[\chi_{1}+\chi_{7}]^*
\, [\chi_{5}+\chi_{11}] + {\rm c.c.}\Big\}}$ \\
& \hspace{0.5cm} $Z_{g,g}= \,{\displaystyle  \; |\chi_{4}+\chi_{8}|^2 - 
\Big\{[\chi_{1}+\chi_{7}]^* \,
[\chi_{5}+\chi_{11}] + {\rm c.c.}\Big\}}$ \\
\hline
\noalign{\medskip}
\hline
 & $S_3$--frustrated $su(2)$ partition functions ($\omega=e^{2 \pi i /3}$) \\
\hline \hline
$\matrix{D_4 \cr \mbox{{\footnotesize $n=6$}}}$ & 
$Z_{e,e} = |\chi_{1} + \chi_{5}|^2 + 2|\chi_{3}|^2$ \qquad 
$Z_{e,a} = |\chi_{1} - \chi_{5}|^2 $ \qquad 
$Z_{e,b} = |\chi_{1} + \chi_{5}|^2 - |\chi_{3}|^2$ \\
 & $Z_{a,e} = |\chi_2 + \chi_4|^2$ \qquad $Z_{a,a} = |\chi_2 - \chi_4|^2$ \\
 & $Z_{b,b^k} = |\chi_{3}|^2 + \omega^k \; (\chi_{1}+\chi_{5})^* \; \chi_{3} +
\omega^{2k} \; \chi_{3}^* \;(\chi_{1} + \chi_{5})$ \\
\hline
\end{tabular}
\caption{\footnotesize List of all frustrated partition functions for the
$\widehat{su}(2)_k$, $n=k+2$, affine theories. The affine characters are labelled by
shifted weights of $P_{++}^{(n)} = \{1 \leq p \leq n-1\}$. For the $D_4$ model, the
partition functions have been labelled by the $S_3$ conjugacy classes $e$, $[a]$ (order
2) and $[b]$ (order 3).}

\end{table}


\medskip
\leftskip=0.8cm
\noindent
\makebox[0pt]
{
\raisebox{-23mm}[0mm][0mm]
{\hspace*{-4mm}\rule{0.2mm}{26mm}}
}%
\noindent
{\sl The $\widehat{su}(2)_k$ models, labelled ADE, have a symmetry group exactly equal
to the automorphism group of the associated Dynkin diagram, namely no symmetry at all
for $E_7$ and $E_8$, the permutation group $S_3$ for $D_4$, and the $\Z_2$ group in all
other cases. When the height $n=k+2$ is odd however, the $\Z_2$ symmetry in the
diagonal models $A_{n-1}$ can only be realized projectively.}

\leftskip=0cm
\medskip \noindent
The corresponding frustrated partition functions are given in Table 1. Let us note that
the theories $D_{n/2+1}$ for $n=0 \bmod 4$, and $E_6$ all have an extended symmetry
algebra and a $\Z_2$ symmetry group. In the $E_6$ model only does the symmetry
group preserve the extended algebra (as easily understood from the conformal embedding
$\widehat{su}(2)_{10} \subset \widehat{sp}(4)_1$, the $\Z_2$ group corresponding to
the $\widehat{sp}(4)$ simple currents).

A consistency check can be made on the results of Table 1 by computing orbifold
partition functions, given by (\ref{orb}). Quotienting by symmetry (sub)groups
realized non--projectively, one finds the following relations
\beq
A_{n-1} \stackrel{\Z_2}{\longleftrightarrow} D_{{n \over 2}+1} \quad \hbox{($n$
even)}, \qquad E_6 \stackrel{\Z_2}{\longleftrightarrow} E_6, \qquad 
D_4 \stackrel{\Z_3}{\longleftrightarrow} D_4, \qquad 
D_4 \stackrel{S_3}{\longrightarrow} A_5.
\label{orb11}
\eeq

\noindent
{\it (b) \underline{su(3) models}}

\bigskip \noindent
In analogy with the notation used for $su(2)$, the $su(3)$ invariants are
named ADE with the height $n=k+3$ as a subscript. Because the charge conjugation
$S^2=C$ commutes with $S$ and $T$, all modular invariants have a partner A$^*$D$^*$E$^*$
obtained by replacing the matrix $M_{p,p'}$ by $M_{p,C(p')}=M_{C(p),p'}$. Four modular 
invariants are however self--conjugate: $D^{}_6=D_6^*$, $D^{}_9=D_9^*$,
$E_{12}^{}=E_{12}^*$ and $E_{24}^{}=E_{24}^*$. These equalities have interesting
consequences on the symmetries and the graphs.

Conjugate theories have always the same symmetry, since the replacement 
\beq
M_{p,p'}^{(g)}(g') \longrightarrow M_{p,C(p')}^{(g)}(g')
\eeq
in all sectors provide a realization of the symmetry in the conjugate theory. One can
readily notice that, if this replacement has no effect in the periodic sector
of self--conjugate theories, it may not be so in the twisted sectors, where it can give
rise to distinct partition functions. As a consequence,  self--conjugate modular
invariant partition functions may be compatible with two different realizations of a
given symmetry group. This indeed happens for the $D_6$ and $D_9$ models. 

Our results for the symmetries of the $su(3)$ affine models are the following. 

\medskip
\leftskip=0.8cm
\noindent
\makebox[0pt]
{
\raisebox{-47mm}[0mm][0mm]
{\hspace*{-4mm}\rule{0.2mm}{50mm}}
}%
\noindent
{\sl All models from the $A_n^{(*)}$, $D_n^{(*)}$ ($n \neq 6,9$) series, as well as the
exceptionals $E^{(*)}_8$ and $E_{12}^{}=E_{12}^*$ have a $\Z_3$ symmetry group;\\
the self--conjugate model $D^{}_6=D_6^*$ has a symmetry group equal to the alternating
group $A_4$, which can however be realized in two different ways;\\
the other self--conjugate model $D^{}_9=D_9^*$ has
$\Z_3$ as symmetry group, also with two different realizations;\\
finally $E_{12}^{\rm MS}$, $E_{12}^{*\,{\rm MS}}$ and $E_{24}^{}$ have no
symmetry at all.

\smallskip \noindent
Moreover the symmetry of the two models $E^{}_8$ and $E^{*}_8$ can be promoted to
$\Z_6$, and that of $E_{24}$ to the group $\Z_2$, provided we allow for projective
representations in twisted sectors.}

\leftskip=0cm
\medskip \noindent
All frustrated partition functions are given in Tables 2 and 3, while Table 5 collects
the projective $\Z_6$ partition functions in the $E^{(*)}_8$ models. Included in Table
4 is a summary of the symmetry groups.

\begin{table}
\hspace{-0.6cm}
\begin{tabular}[h]{|c|l|}
\hline
 & $\Z_3$--frustrated $su(3)$ partition functions ($g^3=e,\, \omega=e^{2 \pi i/3})$\\ 
\hline \hline
$\matrix{A_n \cr \mbox{\scriptsize $n \geq 3$}}$ & 
$Z_{e,g^k}= \displaystyle {\sum_{p} \, \omega^{kt(p)} |\chi_{p}|^2}$ \quad
$Z_{g,g^k} = {\displaystyle {\sum_{p}  \,
\omega^{kt(\mu^{2}(p))}\chi^*_{p}\,\chi^{}_{\mu(p)}}} \quad 
Z_{g^2,g^k} = {\displaystyle {\sum_{p} \,
\omega^{kt(\mu(p))}\chi^*_{p}\,\chi^{}_{\mu^{2}(p)}}}$ \\
\hline
$\matrix{D_n \cr \mbox{\scriptsize$n \neq 0 \bmod 3$}}$ & 
$Z_{e,g^k} = {\displaystyle {\sum_{p}  \, \omega^{kt(p)}
\chi^*_{p}\,\chi^{}_{\mu^{n t(p)}(p)}}}$ \\
\mbox{\scriptsize $(n \geq 5)$} & $Z_{g,g^k} = {\displaystyle {\sum_{t(p) = 2n} 
|\chi_{p}|^2 }} + \omega^{2kn} {\displaystyle {\sum_{t(p) = 0} 
\chi^*_{p}\,\chi^{}_{\mu(p)} + \omega^{kn} \sum_{t(p) 
= n} \chi^*_{p}\,\chi^{}_{\mu^{2}(p)} }}$ \\
& $Z_{g^2,g^k} = {\displaystyle {\sum_{t(p) = n} |\chi_{p}|^2 }} +
{\displaystyle {\omega^{kn} \sum_{t(p) = 0} \chi^*_{p}\chi^{}_{\mu^{2}(p)} + 
\omega^{2kn} \sum_{t(p) = 2n} \chi^*_{p}\chi^{}_{\mu(p)}}}$ \\
\hline
$\matrix{D_n \cr \mbox{\scriptsize $n = 0 \bmod 3$}}$ &  
$Z_{e,g^k} = {\displaystyle \sum_{t(p)=0} \; \bigg[ 
\sum_{j=0}^{2} \; \omega^{kj} \chi^*_{p}\,\chi^{}_{\mu^{j}(p)} \bigg] }$ \\
\mbox{\scriptsize $(n \geq 6)$} & $Z_{g,g^k} = {\displaystyle \sum_{t(p)=2} \;
\bigg[\sum_{j=0}^{2} \; \omega^{kj} \chi^*_{p}\,\chi^{}_{\mu^{j}(p)} \bigg] } 
\qquad \qquad Z_{g^2,g^k} = {\displaystyle \sum_{t(p)=1} \;
\bigg[ \sum_{j=0}^{2} \; \omega^{kj} \chi^*_{p}\,\chi^{}_{\mu^{j}(p)} \bigg] }$ \\
\hline
& $Z_{e,g^k} =  
|\chi^{}_{(1,1)}+\chi^{}_{(3,3)}|^2 + \omega^k |\chi^{}_{(3,1)}+\chi^{}_{(3,4)}|^2 + 
\omega^{2k} |\chi^{}_{(1,3)}+\chi^{}_{(4,3)}|^2$ \\
$\matrix{E_8 \cr \mbox{\footnotesize $n=8$}}$ & 
$\qquad \quad  + |\chi^{}_{(4,1)}+\chi^{}_{(1,4)}|^2 + 
\omega^k |\chi^{}_{(2,3)}+\chi^{}_{(6,1)}|^2
+\omega^{2k}|\chi^{}_{(3,2)}+\chi^{}_{(1,6)}|^2$ \\ 
 & $Z_{g,g^k}= 
(\chi^{*}_{(2,3)}+\chi^{*}_{(6,1)})(\chi^{}_{(1,6)}+\chi^{}_{(3,2)}) +
(\chi^{*}_{(3,1)}+\chi^{*}_{(3,4)})(\chi^{}_{(1,3)}+\chi^{}_{(4,3)})$ \\
 & $ \qquad +
\omega^k \Big[ (\chi^{*}_{(1,6)}+\chi^{*}_{(3,2)})(\chi^{}_{(1,1)}+\chi^{}_{(3,3)}) +
(\chi^{*}_{(1,3)}+\chi^{*}_{(4,3)})(\chi^{}_{(1,4)}+\chi^{}_{(4,1)}) \Big]$ \\
& $ \qquad + \omega^{2k} 
\Big[(\chi^{*}_{(1,1)}+\chi^{*}_{(3,3)})(\chi^{}_{(2,3)}+\chi^{}_{(6,1)}) +
(\chi^{*}_{(1,4)}+\chi^{*}_{(4,1)})(\chi^{}_{(3,1)}+\chi^{}_{(3,4)}) \Big]$ \\
 & $Z_{g^2,g^k} = 
(\chi^{*}_{(1,3)}+\chi^{*}_{(4,3)})(\chi^{}_{(3,1)}+\chi^{}_{(3,4)}) +
(\chi^{*}_{(1,6)}+\chi^{*}_{(3,2)})(\chi^{}_{(2,3)}+\chi^{}_{(6,1)})$ \\
& \qquad  $+ \omega^{2k} \Big[ 
(\chi^{*}_{(2,3)}+\chi^{*}_{(6,1)})(\chi^{}_{(1,1)}+\chi^{}_{(3,3)}) +
(\chi^{*}_{(3,1)}+\chi^{*}_{(3,4)})(\chi^{}_{(1,4)}+\chi^{}_{(4,1)}) \Big]$ \\
& \qquad  $+ \omega^k 
\Big[(\chi^{*}_{(1,1)}+\chi^{*}_{(3,3)})(\chi^{}_{(1,6)}+\chi^{}_{(3,2)}) +
(\chi^{*}_{(1,4)}+\chi^{*}_{(4,1)})(\chi^{}_{(1,3)}+\chi^{}_{(4,3)}) \Big]$ \\
\hline
& $Z_{e,e} = 
|\chi^{}_{(1,1)}+\chi^{}_{(10,1)}+\chi^{}_{(1,10)}+\chi^{}_{(5,2)}+\chi^{}_{(2,5)}+
\chi^{}_{(5,5)}|^2  + 2 |\chi^{}_{(3,3)}+\chi^{}_{(3,6)}+\chi^{}_{(6,3)}|^2$ \\
$\matrix{E_{12} \cr \mbox{\footnotesize $n=12$}}$ & $Z_{e,g^k} = 
|\chi^{}_{(1,1)}+\chi^{}_{(10,1)}+\chi^{}_{(1,10)}+\chi^{}_{(5,2)}+\chi^{}_{(2,5)}+
\chi^{}_{(5,5)}|^2 - |\chi^{}_{(3,3)}+\chi^{}_{(3,6)}+\chi^{}_{(6,3)}|^2 $ \\
& $Z_{g,g^k}=Z_{g^2,g^{-k}} = 
|\chi^{}_{(3,3)}+\chi^{}_{(3,6)}+\chi^{}_{(6,3)}|^2$ \\
& + $\Big[\omega^{k} (\chi^{*}_{(3,3)}+\chi^{*}_{(3,6)}+\chi^{*}_{(6,3)})
(\chi^{}_{(1,1)}+\chi^{}_{(10,1)}+\chi^{}_{(1,10)}+\chi^{}_{(5,2)}+\chi^{}_{(2,5)}+
\chi^{}_{(5,5)}) + {\rm c.c.} \Big]$ \\
\hline
\end{tabular}
\caption{\footnotesize List of frustrated partition functions for affine
$\widehat{su}(3)_k$ ($n=k+3$) theories. The characters are labelled by shifted weights
of $P_{++}^{(n)} = \{p=(a,b) : a, b \geq 1, a+b \leq n-1\}$. The automorphism of
$P_{++}^{(n)}$ is $\mu(a,b)=(n-a-b,a)$ and $t(a,b)=a+2b\,\bmod 3$ is the triality. The
partition functions of the conjugate theories are related to the above ones via the
action of the conjugation operator $C(a,b) = (b,a)$.}
\end{table}


\begin{table}[th]
\begin{center}
\begin{tabular}{|c|l|}
\hline
& $A_4$--frustrated $\widehat{su}(3)_3$ partition functions \\
\hline
\hline
&  $Z_{e,e}= 
|\chi_{(1,1)}+\chi_{(1,4)}+\chi_{(4,1)}|^2  + 3 \, |\chi_{(2,2)}|^2$ \\
$\matrix{D^{}_6 \cr \mbox{\footnotesize $n=6$}}$ & 
\hspace{5mm} $Z_{e,a} = |\chi_{(1,1)}+\chi_{(1,4)}+\chi_{(4,1)}|^2 - 
|\chi_{(2,2)}|^2$ \\
& \hspace{5mm} $Z_{e,b^k} = |\chi_{(1,1)} + \omega^k \, \chi_{(1,4)} + 
\omega^{2k} \, \chi_{(4,1)}|^2 \qquad (k=1,2)$ \\
& $Z_{a,e} = 2\,|\chi_{(2,2)}|^2 + 
\Big[\chi^*_{(2,2)}\,(\chi_{(1,1)}^{}+\chi_{(1,4)}^{}+\chi_{(4,1)}^{}) 
+ {\rm c.c.} \Big]$ \\
& \hspace{5mm} $Z_{a,a} = 2\,|\chi_{(2,2)}|^2 - 
\Big[\chi^*_{(2,2)}\,(\chi_{(1,1)}^{}+\chi_{(1,4)}^{}+\chi_{(4,1)}^{}) 
+ {\rm c.c.} \Big]$ \\
& \hspace{5mm} $Z_{a,a'} = \chi^*_{(2,2)}\,(\chi_{(1,1)}^{}+\chi_{(1,4)}^{}
+ \chi_{(4,1)}^{})  - {\rm c.c.} $ \\
& \hspace{5mm} $Z_{a,aa'} = -\chi^*_{(2,2)}\,(\chi_{(1,1)}^{}+\chi_{(1,4)}^{}
+ \chi_{(4,1)}^{})  - {\rm c.c.} = -Z_{a,a'}$ \\
& $Z_{b,e} = |\chi_{(1,2)}+\chi_{(2,3)}+\chi_{(3,1)}|^2$ \\
& \hspace{5mm} $Z_{b,b^k} = |\chi_{(1,2)} + \omega^k \, \chi_{(2,3)} + 
\omega^{2k} \, \chi_{(3,1)}|^2 \qquad (k=1,2)$ \\
& $Z_{b^2,e} = |\chi_{(2,1)}+\chi_{(3,2)}+\chi_{(1,3)}|^2$ \\
& \hspace{5mm} $Z_{b^2,b^k} = |\chi_{(2,1)} + \omega^{2k} \, \chi_{(2,3)} + 
\omega^k \, \chi_{(1,3)}|^2 \qquad (k=1,2)$ \\
& {\rm The second realization $D_6^*$ is obtained from above by the insertion of $C$} \\
\hline
\end{tabular}
\end{center}
\caption{\footnotesize List of $A_4$ partition functions for the $D_6^{}=D_6^*$ model of
$\widehat{su}(3)_3$ (height $n=6$). Characters are labelled by shifted weights, and
partition functions by the four classes of $A_4$: in the standard notation of
permutations in terms of cycles, $[a]=(\cdot\cdot)(\cdot\cdot)$, and the two classes
$[b],[b^2]$ correspond to $(\cdot\cdot\cdot)(\cdot)$. The centralizer of the class
$[a]$ is $C(a)=\langle a,a' \rangle \sim \Z_2 \times \Z_2$ which contains all elements
of order 2 and the identity. The centralizers of the other two classes are $C(b) =
C(b^2) = \langle b \rangle$, isomorphic to $\Z_3$.}
\end{table}


As explained above, the two realizations of the symmetry in the $D^{(*)}_6$ and
$D^{(*)}_9$ models are related to each other by the action of $C$. Concretely, the
difference manifests itelf in the sectors twisted by order 3 group elements, meaning,
in both cases, that the $\Z_3$ possesses two different realizations, say
$\Z_3$ and $\Z_3'$. One easily checks that these two realizations are not compatible
with each other, therefore excluding a symmetry $\Z_3 \times \Z'_3$. So, for the
$D^{(*)}_9$ models, the symmetry group $G$ should be non--abelian and have two
non--conjugate $\Z_3$ subgroups (and only these), whereas for the $D^{(*)}_6$ models,
$G$ must in addition have one $\Z_2 \times \Z_2$ subgroup. It is not difficult to see,
in either case, that such a non--abelian group does not exist (the argument is the same
in both cases and is given in Section 3.3.1), and this means that $\Z_3$ and $\Z_3'$ are
parts of two different groups. So the modular invariant partition function $D^{(*)}_9$
is compatible with two realizations of a cyclic $\Z_3$ group, while $D^{(*)}_6$ is
compatible with two realizations of a $A_4$ group ---the only one to contain one $\Z_3$
and one $\Z_2 \times \Z_2$, up to conjugation, and nothing else.

This strongly suggests that there are two pairs of models $(D^{}_6, D^{*}_6)$ and
$(D^{}_9, D^{*}_9)$ rather than two models. The models within each pair share the
same periodic partition function on the torus and the same symmetry group, but 
differ in the field content of their $\Z_3$--twisted sectors. The same suggestions,
based on the existence of isospectral graphs, have been made in \cite{bppz}.

With the help of Tables 2 and 3, one can establish that the orbifold procedure yields
the following mappings between the various models (we mean orbifold with respect to the
symmetry (sub)groups which are realized non--projectively):
\bea
& A_n \stackrel{\Z_3}{\longleftrightarrow} D_n \qquad  \quad
A^*_n \stackrel{\Z_3}{\longleftrightarrow} D^*_n \qquad \hbox{(all $n$)} \label{orb21}
\\
\noalign{\smallskip}
& D^{(*)}_6 \stackrel{\Z_2, \, \Z_2 \times \Z_2}{\longleftrightarrow} D^{(*)}_6 \qquad 
\quad  D^{}_6 \stackrel{A_4}{\longrightarrow} A^{}_6 \qquad  \quad 
D^{*}_6 \stackrel{A_4}{\longrightarrow} A^{*}_6 \label{orb22} \\
\noalign{\smallskip}
& E^{}_8 \stackrel{\Z_3}{\longleftrightarrow} E^*_8 \qquad  \quad
E^{}_{12} \stackrel{\Z_3}{\longleftrightarrow} E^{}_{12}.
\label{orb23} 
\eea

\subsection{Comparison with graphs}

There is a number of issues that one can address if one views the above results in a
graph theoretic light. We will specifically focus on three issues: the symmetry group
itself, the action of $G$ in the various sectors and the orbifold procedure. 

The most basic question concerns the identification of the symmetry group of an affine
model with the automorphism group of its graph(s). The answer is mostly positive for the
$su(2)$ models: if one insists on non--projective realizations, the symmetry of the
field theory is exactly the automorphism group of its (Dynkin) graph except in the
infinite series of odd level diagonal theories $A_{n-1}$ ($n$ odd) which have no
symmetry (though the graph has a $\Z_2$ automorphism). If we allow projective
realizations, the identification of the two groups holds for all theories. This
difference in the odd level diagonal theories leave a trace in their minimal models,
as it was already observed in \cite{r} that the conformal models based on a pair of
A--algebras, both of even rank, do not have a non--projective realization of the
($\Z_2$) symmetry of their graphs. 

The situation for the $su(3)$ theories, summarized in Table 4, is not as neat. One can
see that, in the majority of cases, the symmetry of the graph does not coincide with
that of the corresponding theory. That statement must be qualified for the two series
${\cal A}^*_n$ and ${\cal D}^{}_{n \neq 0(3)}$, and also for the isolated model
${\cal E}^*_8$, since they are quotients of graphs by a symmetry group that acts freely
(fold graphs). In such  cases one naturally expects that this explicit symmetry
disappear (while some other can emerge). In all other cases (the 3--colourable graphs),
the symmetry group $G$ of the field theory is a subgroup of the automorphism group of
its graph(s), of index 1,  2, 4 or even 6 in one case. The situation does not improve
much by allowing projective realizations since only in the $E_8^{(*)}$ and $E^{}_{24}$
models does the symmetry group get enlarged (from $\Z_3$ to $\Z_6$, or from $\{e\}$ to
$\Z_2$). 

On the question of how $G$ acts on the fields, the answer is universal and
strengthens the connection with the graphs recalled in Section 2.4 (see \cite{r} for
similar statements in the minimal models). Whenever $G \subset {\rm Aut}\,\Gamma$
(i.e. all graphs for $su(2)$ and all 3--colourable ones for $su(3)$), it turns out that,
in addition to encoding the diagonal terms of the periodic partition function, the graph
also specifies the characters of the representations by which $G$ acts in the periodic
sector. Indeed $G$ has two different actions. On the field theoretic side, it acts on
the affine representations labelled $(p,p)$ that occur in the periodic Hilbert space. On
the graph theoretic side, being a subgroup of Aut$\,\Gamma$, it acts on the eigenspaces
$V_i$ of the adjacency matrix $n_f$ of the graph, by some representation $R_i$. Then the
fact is that the two actions coincide (their characters are equal): for all $g \in G
\subset {\rm Aut}\,\Gamma$, one has
\beq
\left\{\matrix{{}^gV_i = R_i(g) V_i \cr 
\noalign{\smallskip} {\rm Tr}\,R_i(g) = \lambda_i(g) }\right\}
\qquad \Longleftrightarrow
\qquad Z_{e,g} = \sum_{i \in {\cal E}} \lambda_i(g)\,|\chi_i|^2 + \hbox{non--diagonal}.
\label{corr}
\eeq
This extends to $g \neq e$ the correspondence recalled in Section 2.4 between the
graphs and the modular invariant partition functions. It should be stressed however
that the correspondence (\ref{corr}) for $g \neq e$ appears as a bonus, since it was
not put in from the start, contrary to the case $g = e$.

To consider and investigate (\ref{corr}) for the graph automorphisms $g$ other than
those in $G$ is natural, as it yields sensible functions $Z_{e,g}$, from which one can
start. For a reason that is not clear to us, they always lead to functions $Z_{g,e}$
with non--integer coefficients, except in four cases, the odd level diagonal $su(2)$
models and the $E^{}_8,\,E^*_8,\,E^{}_{24}$ models of $su(3)$. For these, the graphs
have a $\Z_2$ symmetry which yields through (\ref{corr}) sensible functions $Z_{e,g}$
and $Z_{g,e}$, but produces $\pm i$ coefficients (charges) in $Z_{g,g}$. One can easily
see that these three functions are fully compatible with a projective action of that
$\Z_2$ in the twisted sector, in the sense of Section 2.2. In the case of the $E_8$
model, this projective $\Z_2$ combines with a $\Z_3$ to form a projective
$\Z_6$, see Section 3.3.4 for more details.   

\begin{table}
\hspace{-0.6cm}
\begin{center}
\mbox{\footnotesize
\begin{tabular}[t]{|c||c|c|c|c|c|c|c|c|c|c|}
\hline
$\Gamma$ & ${\cal A}^{}_n$ & ${\cal A}^*_{4,2m+1}$ & ${\cal A}^*_{2m \geq 6}$ & ${\cal
D}^{}_6$ & ${\cal D}^{}_{3m}$ & ${\cal D}^{}_{3m\pm 1}$ & ${\cal D}^*_6$ & ${\cal
D}^*_{n\geq 7}$ & ${\cal E}^{}_8$ & ${\cal E}^*_8$ \\ 
\hline
\hline
${\rm Aut}\,\Gamma$ & $\Z_3$ & $-$ & $\Z_2$ & $S_4$ & $S_3$ & $-$ & $A_4 \times \Z_2$ &
$\Z_3$ & $\Z_6$ & $\Z_2$ \\
\hline
$G$ & $\Z_3$ & $\Z_3$ & $\Z_3$ & $A_4$ & $\Z_3$ & $\Z_3$ & $A_4$ & $\Z_3$ & $\Z_3$ &
$\Z_3$ \\
\hline
$G_{\rm proj}$ &  &  &  &  &  &  &  &  & $\Z_6$ & $\Z_6$ \\
\hline
\end{tabular}}
\end{center}
\begin{center}
\mbox{\footnotesize
\begin{tabular}[t]{|c||c|c|c|c|c|c|}
\hline
cont'd & ${\cal E}^{(1)}_{12}$ & ${\cal E}^{(2)}_{12}$ & ${\cal E}^{(3)}_{12}$ & ${\cal
E}^{(4)}_{12}$ & ${\cal E}^{(5)}_{12}$ & ${\cal E}^{}_{24}$ \\ 
\hline
\hline
& $S_3$ & $S_3$ & $S_3 \times \Z_3$ & $\Z_2 \times \Z_2$ & $-$ & $\Z_2$ \\
\hline
& $\Z_3$ & $\Z_3$ & $\Z_3$ & $-$ & $-$ & $-$ \\
\hline
&  &  &  &  &  & $\Z_2$ \\
\hline
\end{tabular}}
\end{center}
\caption{\footnotesize Synopsis of the groups of symmetry pertaining to the $su(3)$
affine models and their graphs, in a notation borrowed from \cite{bppz}. The top line
designates the graphs: the superscript $i$ appended to the graph ${\cal E}^{}_{12}$
labels three isospectral graph, ${\cal E}^{(4)}_{12}$ and ${\cal E}^{(5)}_{12}$
correspond respectively to the invariants $E_{12}^{*\,{\rm MS}}$ and $E_{12}^{\rm
MS}$. The second line gives the automorphism group ${\rm Aut}\,\Gamma$ of the various
graphs, whereas the third one mentions the group of symmetry $G$ we found in the field
theories. The last line refers to the group ${\rm Aut}\,\Gamma \supset G_{\rm proj}
\supset G$ of which the corresponding field theory carries projective representations.}
\end{table}

Thus concretely, (\ref{corr}) means that much of the Tables 1,2,3 and 5 can actually be
read off from the graphs, since a spectral analysis of their adjacency matrices yield
explicit expressions for the numbers $M_{p,p}^{(e)}(g)$. When there are several
isospectral graphs for a given modular invariant partition function, the numbers
$M_{p,p}^{(e)}(g)$ one gets may or may not differ from graph to graph. For the $su(3)$
invariants $D^{(*)}_6$ or $D^{(*)}_9$, there are two isospectral graphs ${\cal D}^{}_6$,
${\cal D}^*_6$ and ${\cal D}^{}_9$, ${\cal D}^*_9$ which have a different automorphism 
group, and which yield different values for $M_{p,p}^{(e)}(g)$, which in turn lead to
two different realizations of the symmetry. In contrast, there are three graphs ${\cal
E}^{(i)}_{12}$ associated to the single invariant $E_{12}^{}$, but all three have a
$\Z_3$ automorphism subgroup which gives the same numbers $M_{p,p}^{(e)}(g)$.

The correspondence (\ref{corr}) has a counterpart in the non--periodic, twisted sectors.
For the same class of models, such that $G \subset {\rm Aut}\,\Gamma$, the diagonal
terms of $Z_{g,e}$ are encoded in the subgraph of $\Gamma$ made of the nodes that are
fixed by $g$, via the same relation as before. Namely, if $\Gamma^{(g)}$ denotes the
subgraph of $\Gamma$ fixed by $g$ (possibly empty), with adjacency matrix $n_f^{(g)}$,
it appears in most cases that
\beq
{\rm spec}\,n_f^{(g)} = \{{S_{f,j} \over S_{0,j}} \;:\; j \in {\cal E}_g\} \qquad
\Longleftrightarrow \qquad Z_{g,e} = \sum_{i \in {\cal E}_g} \; |\chi_i(q)|^2 + 
\hbox{non--diagonal}.
\label{fixed}
\eeq
The only cases for which the connection breaks down are the graphs which contain
multiple (double) links, namely the ${\cal D}_{3n}$ series and the graph ${\cal
E}_{12}^{(1)}$, all in $su(3)$. All of them have a $\Z_3$ symmetry, with a fixed point
graph which has the correct number of nodes but not the correct eigenvalues. For the
graphs ${\cal D}_{3n}$, the sets ${\cal E}_g$ and ${\cal E}_{g^2}$ are such that
$\{S_{f,j}/S_{0,j}\}_{j \in {\cal E}_g \; {\rm or} \; {\cal E}_{g^2}}$ are not even
closed under complex conjugation, and so cannot possibly form the spectrum of an
integer matrix. 

As to the way $G$ (or a proper centralizer) acts in a twisted sector ${\cal
H}_g$, in all but one case, one can see from the tables that all diagonal fields in all
twisted sectors are invariant under the relevant symmetry group (centralizer). This is
again consistent, since the subgraph $\Gamma^{(g)}$ has no symmetries left over from
$G$. The only exception is in the $\Z_2$--twisted sector of the $D_6^{}$, $D_6^*$
models, see Table 3. The same observation has been made in the minimal conformal
models, and was actually a happy fact in the analysis of the corresponding boundary
theories, as the nodes of the fixed point graph were put in one--to--one relation with
$G$--invariant boundary conditions \cite{r}. An analogous correspondence is to be
expected in the affine models. 

The last point of comparison we want to make concerns the orbifold procedure. The
results regarding the orbifold partition functions are given in (\ref{orb11}) for
$su(2)$, and in (\ref{orb21}) to (\ref{orb23}) for $su(3)$. As cosets of graphs by
subgroups of their automorphism groups can be defined, the question naturally arises to
see if, starting from a theory $\cal T$ and graph $\Gamma$, the orbifold theory ${\cal
T}/H$ has a graph that is given by the quotient graph $\Gamma/H$ (again we restrict to
cases where $H \subset G \subset {\rm Aut}\,\Gamma$ is realized non--projectively). 

That question is perhaps less significant because it depends on the prescriptions
used to define the quotient graph, especially when the subgroup $H$ has fixed points
and when the graph has multiple links. However in the simplest cases, when the graph has
no multiple links and when the elements of $H$ by which we want to quotient have the
same fixed points, standard prescriptions \cite{k} require to multiplicate all fixed
points before proceeding to the quotient. When there are multiple links, natural though
ad hoc prescriptions can be given as well regarding the way these links must be split.

With the same restrictions as above regarding the non 3--colourable graphs of $su(3)$,
one can check, using these prescriptions, that indeed the quotient graph is the graph
of the orbifold theory (obvious for some classes of graphs, constructed by quotient).
Interestingly, the conjugate graphs ${\cal D}_6^{}$ and ${\cal D}_6^*$ are the orbifold
of each other under a quotient by $\Z_2 \subset A_4$, something that cannot be revealed
at the level of the partition functions because the two theories $D_6^{}$ and $D_6^*$
only differ in their $\Z_3$--twisted sectors. Also, the $\Z_3$ quotient exchanges the
$su(3)$ graph ${\cal E}_{12}^{(1)}$ and ${\cal E}_{12}^{(2)}$, while ${\cal
E}_{12}^{(3)}$ is a self--orbifold. 

The conclusion one should draw from this comparison is that symmetry considerations
makes the relationship between the modular invariants and the graphs tighter. The
$su(2)$ theories are particularly well--behaved in this respect since the connection
for them is complete (exception made of the projective realizations): the graphs give
the correct symmetries, the correct twisted partition functions and the correct
orbifold relations. To a large extent, the same is true of the $su(3)$ models, since
again the graphs are capable to predict much of the symmetries of the corresponding
models. But at the same time, the same considerations make this connection fade, as the
graphs in many cases would predict more symmetries than what is actually realized.
Against this, one could argue that the list of graphs is not exhaustive in $su(3)$ and
that yet to be discovered new graphs would restore a perfect connection. Unlikely
as it is, one cannot hope this to be strictly true: the two isospectral graphs ${\cal
D}_6^{}$ and ${\cal D}_6^*$ are the only ones whose spectrum matches the diagonal
pieces of the $D_6^{(*)}$ modular invariant, and yet both have a bigger symmetry than
$A_4$, the maximal symmetry group that can be realized in the field theory(ies). 

We finish the presentation of our results by turning to some of the most illustrative
and instructive cases, where much of the above results can be seen explicitely.


\subsection{Selecta}

We give in this section a more detailed study of a few particular cases, worth of being
singled out for their peculiarities. Not surprisingly, they all belong to the $su(3)$
graphs. 

\subsubsection{The $D_6^{(*)}$ invariant of su(3)}

This is by far the richest and most interesting case, by many aspects, the most obvious
one being the size of its symmetry group.

Let us recall that this self--conjugate modular invariant, 
\beq
Z(D_6^{},D_6^*) = |\chi_{(1,1)}+\chi_{(1,4)}+\chi_{(4,1)}|^2  + 3 \, |\chi_{(2,2)}|^2,
\label{d6} 
\eeq
has six diagonal terms $|\chi_i|^2$ with ratios $S_{f,i}/S_{0,i} =
2,2\omega,2\omega^2,0,0,0$. These numbers are the eigenvalues of the adjacency matrix
of two (and only two) graphs, noted ${\cal D}_6^{}$ and ${\cal D}_6^*$ (after
\cite{bppz}) and shown in Figure 1.

\begin{figure}[htb]
\leavevmode
\begin{center}
\mbox{\epsfscale 600 \epsfbox{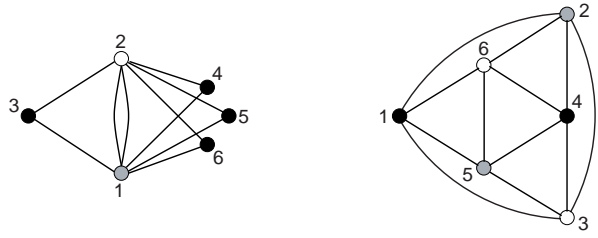}}
\end{center}
\vspace{-1.5cm}
\caption{\footnotesize The isospectral graphs ${\cal D}_6^{}$ (left) and ${\cal D}_6^*$
(right). All links are oriented, f.i. from black to grey to white to black.}
\end{figure}

The automorphism group of ${\cal D}_6^{}$ is $S_4$ and corresponds to all permutations
of the nodes 3 to 6. The three eigenvectors of non--degenerate eigenvalue $2\omega^k$
are invariant under $S_4$, while the other three transform in an irreducible
representation (with character equal to $-1$ on order 4 group elements). Restricted to
$A_4$, it remains irreducible of degree 3, and has character $\lambda_3(g) = (3,-1,0,0)$ for
respectively $g=e$, $g \in [a]$ (the class of order 2 elements), $g \in [b]$ (first
class of order 3 elements) and $g \in [b^2]$ (second class of order 3 elements). Thus,
under the $A_4$ subgroup of Aut$\,{\cal D}_6^{}$, the six eigenvectors of the adjacency
matrix transform in $r_0 \oplus r_0 \oplus r_0 \oplus r_3$, with $r_0$ the trivial
representation. 

The other graph ${\cal D}_6^*$ has automorphism group $A_4 \times \Z_2$ (the graph is
an (oriented) octaedron with base formed f.i. by the nodes 1,3,4,6). The group is
generated by the order 3 rotations and the three commuting transpositions $\sigma_i$
which exchange the nodes $i$ and $i+3$. The combined $\sigma_1 \sigma_2 \sigma_3$ (the
inversion of the octaedron through the origin) generates the center $\Z_2$ of the
group. The $A_4$ subgroup is generated by the rotations and the products $\sigma_i
\sigma_j$. Under the action of $A_4$, the three non--degenerate eigenvectors transform
in the three inequivalent one--dimensonal representations $r_k$, $k=0,1,2$, with
characters $\lambda_k(g)=(1,1,\omega^k,\omega^{2k})$ in the same notation as above,
while the three degenerate eigenvectors transform again in the degree 3 irreducible
representation with character $\lambda_3(g) = (3,-1,0,0)$. Thus the six eigenvectors
transform under $A_4$ as $r_0 \oplus r_1 \oplus r_2 \oplus r_3$.

Starting from the partition function (\ref{d6}), one first checks that it is compatible
with the cyclic symmetries $\Z_2$ (in a unique way in the periodic sector) and $\Z_3$
(in two different ways). The $\Z_2$ partition functions one finds are those called
$Z_{e,a},Z_{a,e},Z_{a,a}$ in Table 3, from which one sees that the periodic sector has
a field three times degenerate, on which the $\Z_2$ generator acts by a representation
of character equal to $-1$. As explained in Section 2.2, these are precisely the
circumstances under which the $\Z_2$ group can extend. Writing $A,B,C$ for
$Z_{e,a},Z_{a,e},Z_{a,a}$, we look for a signed partition function $D$, such that the
following 4--by--4 table makes a consistent set of partitions functions:

\begin{center}
\begin{tabular}[t]{|c||c|c|c|c|c|}
\hline
$Z_{g,g'}$ &  $e$ & $a$ & $a'$ & $ aa'$ \\
\hline \hline
$e$ & $Z_{e,e}$ & $A$ & $A$ & $A$  \\
\hline
$a$ & $B$ & $C$ & $D$ & $T^\dag D T$ \\
\hline
$a'$ & $B$ & $S^\dag D S$ & $C$ & $(ST)^\dag D (ST)$ \\
\hline
$aa'$ & $B$ & $(TS)^\dag D (TS)$ & $(STS)^\dag D (STS)$ & $C$  \\
\hline
\end{tabular}
\end{center}

\smallskip\noindent
Much of it is fixed by the requirement that its restriction to any $\Z_2$
subgroup yields back the known 2--by--2 table. It turns out that there is a
unique\footnote{Up to an overall sign, but since $T^\dag D T = -D$, it reflects the
ambiguity in the choice of the generator of the second $\Z_2$ factor ($a'$ or $aa'$).} 
solution for $D=Z_{a,a'}$, given in Table 3. Thus the group $\Z_2$ indeed extends to
$\Z_2 \times \Z_2$, and to nothing bigger. 

So the partition function (\ref{d6}) is compatible with one $\Z_2 \times \Z_2$
and two $\Z_3$, with corresponding partition functions given in Table 3 (the two $Z_3$
realizations are related via the action of $C$). One easily checks that none of them is
compatible with another one, making them non--commuting subgroups of a bigger
$G$. The argument that such a $G$ does not exist is not difficult, and relies on Sylow's
theorems.

The two $\Z_3$ subgroups are $3$--groups. Either they are Sylow subgroups, which is
impossible since they would be conjugate, contradicting the fact that they realized
differently on the periodic Hilbert space. Or the order of $G$ is divisible by 9, in
which case there is a $3$--subgroup of order 9, which can only be a cyclic $\Z_9$ or a
product $\Z_3 \times \Z_3$. Both possibilities are to be ruled out.

So the two $\Z_3$ subgroups cannot be accomodated within a single group. Retaining only
one of the two leads to the group $G=A_4$, which is therefore the maximal symmetry
group. It can be realized in two different ways through the choice of its $\Z_3$
subgroup, the two realizations being related by conjugation. Moreover, the two
realizations one obtains correspond exactly to the 
$A_4$ action in the two graphs ${\cal D}_6^{}$ and ${\cal D}_6^*$, explicited above,
since they yield respectively (see Table 3)
\beq
Z_{e,g}({\cal D}_6^{}) = |\chi_{(1,1)}|^2 + |\chi_{(1,4)}|^2 + |\chi_{(4,1)}|^2 +
\lambda_3(g)\,|\chi_{(2,2)}|^2 + \hbox{non--diagonal},
\eeq
and
\beq
Z_{e,g}({\cal D}_6^*) = |\chi_{(1,1)}|^2 + \lambda_1(g)\,|\chi_{(1,4)}|^2 +
\lambda_2(g)\,|\chi_{(4,1)}|^2 + \lambda_3(g)\,|\chi_{(2,2)}|^2 + \hbox{non--diagonal},
\eeq
in terms of $A_4$ irreducible characters. These facts provide the basis for our earlier
suggestion that the partition function (\ref{d6}) corresponds to two different models. 

We also note that the (fixed points of the) graph gives the correct number of diagonal
terms in the frustrated partition functions: 3 or 0 in $\Z_3$--twisted sector and 2 in
the $\Z_2$--twisted one.

Finally, the orbifolds, both at the level of the field theories and at the level of the
graphs, have been discussed earlier. The last point we want to comment on concerns the
possibility of making a twisted orbifold, through discrete torsion. The procedure has
been recalled in Section 2.3.

The introduction of discrete torsion requires a (cohomologically) non--trivial
2--cocycle $\omega$ on the orbifold group. As $H^2(A_4,U(1))=\Z_2$, there is a unique
choice for $\omega$. To this $\omega$ corresponds the central extension (double
covering) ${\rm SL}_2(\F_3)$ of $A_4 \sim {\rm PSL}_2(\F_3)$. The character table 
of ${\rm SL}_2(\F_3)$ (tabulated f.i. in \cite{tw}) provides the projective and
non--projective characters of $A_4$. One can see that $A_4$ has three
projective representations, all of degree 2, with zero characters on the class $[a]$.
So only for the three elements $a,a',aa'$ of this class can the quantity
$\epsilon(g,g')$ be different from 1\footnote{The theory of projective representations
\cite{karp} says that the elements $g$ of all other classes are $\omega$--regular,
meaning precisely that $\omega(g,g')=\omega(g',g)$ for all $g' \in C(g)$.}. It can be
most conveniently computed from one explicit representation $R$, since for commuting
elements, one has $R(g)R(g') = \epsilon(g,g') R(g')R(g)$. Choosing any two--dimensional
representation of ${\rm SL}_2(\F_3)$ given in \cite{tw}, one finds
\beq
\epsilon(g,g') = \cases{
+1 & if $g,g' \not\in \{a,a',aa'\}$ or if $g=g'$,\cr
-1 & if $g\neq g' \in \{a,a',aa'\}$.}
\eeq
Thus in the sum (\ref{torb}) giving the orbifold partition function, the part that
concerns the sector ${\cal H}_a$ reads $Z_{a,e} + Z_{a,a} - Z_{a,a'} - Z_{a,aa'}$
whereas the usual orbifold summation would take an all plus combination. This makes
however no difference since $Z_{a,a'} = -Z_{a,aa'}$. So with discrete torsion or not,
the orbifold of the $D_6^{}$ (resp. $D_6^*$) model by its $A_4$ symmetry is the model
$A_6^{}$ (resp. $A_6^*$) with symmetry $\Z_3$, equal, as expected, to the quotient of
$A_4$ by its commutator subgroup $\Z_2 \times \Z_2$.

\subsubsection{The $D_9^{(*)}$ invariant of su(3)}

This case is similar to the previous case except that the symmetry is smaller. All that
has been said for the two $\Z_3$ realizations in the $D_6^{(*)}$ can be repeated here.
In particular there is no group $G$ that can accomodate them both, so that the same
argument points to the existence of two separate field theories, $D_9^{}$ and
$D_9^{*}$, with the same torus partition function. 

\subsubsection{The $E_{12}^{}$ invariant of su(3)}

This is the last of the three cases where several isospectral graphs, here three, are
known to correspond to the same modular invariant. The three graphs, noted ${\cal
E}_{12}^{(i)}$ in \cite{bppz}, are shown in Figure 2.

\begin{figure}[htb]
\leavevmode
\begin{center}
\mbox{\epsfscale 600 \epsfbox{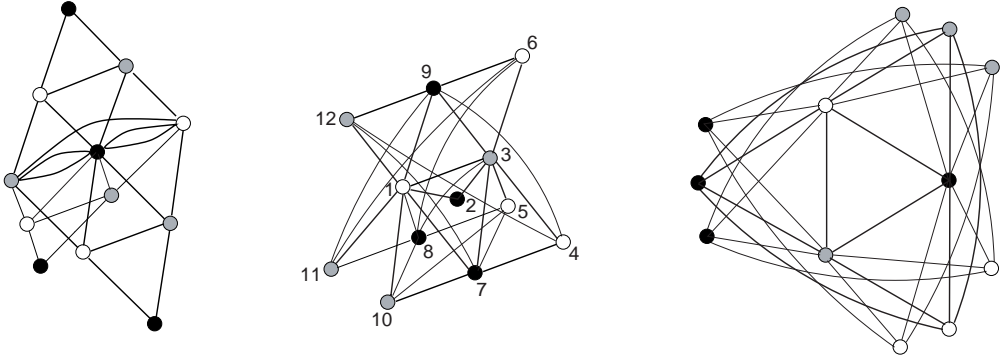}}
\end{center}
\caption{\footnotesize The three isospectral graphs ${\cal E}_{12}^{(1)}$ (left),
${\cal E}_{12}^{(2)}$ (middle) and ${\cal E}_{12}^{(3)}$ (right), corresponding to
the self--conjugate modular invariant $E_{12}$.}  
\end{figure}

The field theory has a maximal $\Z_3$ symmetry, with partition functions (see Table 2)
\beq
Z_{e,g^k} = |\chi^{}_{(1,1)} + \chi^{}_{(10,1)} + \chi^{}_{(1,10)} + \chi^{}_{(5,2)} +
\chi^{}_{(2,5)} + \chi^{}_{(5,5)}|^2 + (\omega^k + \omega^{2k}) \,
|\chi^{}_{(3,3)}+\chi^{}_{(3,6)}+\chi^{}_{(6,3)}|^2.
\label{way}
\eeq
with $\omega$ a primitive third root of 1. It is a self--orbifold theory. One would
like to see if these facts, compared with the graph data, can help select one of the
three graphs. 

Their automorphism group is $S_3$, $S_3$ and $S_3 \times \Z_3$ respectively. In
${\cal E}_{12}^{(1)}$, the $S_3$ simply permutes the three wings attached to the
central axis. In ${\cal E}_{12}^{(2)}$, the order 3 automorphisms are rotations
around the axis; the order 2 elements are the conjugates of $(4 \leftrightarrow 6, 
7 \leftrightarrow 8, 11 \leftrightarrow 12)$. In ${\cal E}_{12}^{(3)}$, the
factor $\Z_3$ are the rigid rotations of the graph, and the $S_3$ permutes the three
nodes of each peripheral group, the same way within each group. 

In each case, the unique, up to conjugation, $\Z_3$ subgroup of $S_3$ acts on the
eigenvectors of their adjacency matrix in the way shown by the diagonal terms of
(\ref{way}). The third graph ${\cal E}_{12}^{(3)}$ is its own orbifold under the $\Z_3$
subgroup of $S_3$, whereas the first two are the orbifold of each other, provided the
double links of ${\cal E}_{12}^{(1)}$ are handled properly. The fixed point graphs are
all equal to an oriented triangle (again with an ad hoc prescription for the double
links), whose adjacency matrix has the three third roots of 1 as eigenvalues. These are
also the values of the ratios $S_{f,i}/S_{0,i}$ for the three diagonal terms in the
twisted sectors, $i=(3,3),\,(3,6),\,(6,3)$. Thus the only feature that distinguishes
them is the fact the third one is its own orbifold. Esthetically, this property may seem
desirable as the field theory is its own orbifold too, but the argument is not
compelling. 

Other methods \cite{pz,xu,bek} point to ${\cal E}_{12}^{(1)}$ as the graph that is
genuinely associated with the $E_{12}^{}$ theory.

\subsubsection{The $E_8^{}$ invariant of su(3)}

The graph corresponding to the $\widehat{su}(3)_5$ invariant $E_8^{}$ is shown in
Figure 3. Its automorphism group $\Z_6$ acts by rigid rotations. 

\begin{figure}[htb]
\leavevmode
\begin{center}
\mbox{\epsfscale 500 \epsfbox{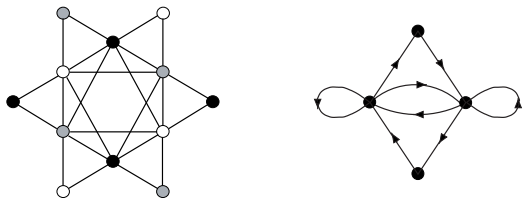}}
\end{center}
\caption{\footnotesize The graph ${\cal E}_8^{}$ (left) corresponding to the level 5
$su(3)$ modular invariant $E_8^{}$, and the conjugate graph ${\cal E}^*_8$ (right).}  
\end{figure}

Its $\Z_3$ subgroup is the maximal symmetry that has a non--projective realization in
the field theory, and leads to the partition functions displayed in Table 2. As to the
$\Z_2$ subgroup, the graph data and the relation (\ref{corr}) allow to write a unique
$Z_{e,g}$ which yields a sensible integer function $Z_{g,e}(\tau) = Z_{e,g}({-1 \over
\tau})$. The $T$ transform of the latter, usually equal to $Z_{g,g}$, acts on the
various fields with phases $\pm i$, which indeed suggest a projective action. Combined
with the $\Z_3$, one expects a projective action of $\Z_6$. Let us show that it is
indeed the case.

Let $g$ be a generator of $\Z_6$. The graph data provide a specific form 
for the diagonal terms of $Z_{e,g}$, which one tries to complete so as to get a integer
function $Z_{g,e}(\tau) = Z_{e,g}({-1 \over \tau})$. As before, this fixes it uniquely
to (with $\zeta$ a primitive twelveth root of 1)
\bea
Z_{e,g} &=& |\chi^{}_{(1,1)}+\chi^{}_{(3,3)}|^2 
+ \zeta^2 \, |\chi^{}_{(1,3)}+\chi^{}_{(4,3)}|^2
+ \zeta^{4} \, |\chi^{}_{(2,3)}+\chi^{}_{(6,1)}|^2
+ \zeta^{6} \, |\chi^{}_{(1,4)}+\chi^{}_{(4,1)}|^2 \nonumber \\
&& \hskip 1truecm  
+ \zeta^{8} \, |\chi^{}_{(1,6)}+\chi^{}_{(3,2)}|^2 
+ \zeta^{10} \, |\chi^{}_{(3,1)}+\chi^{}_{(3,4)}|^2.
\label{zeg}
\eea


\begin{table}[t]
\begin{center}
\begin{tabular}{|c|l|}
\hline
$n=8$ & Projective $\Z_6$-frustrated $\widehat{su}(3)_5$ partition functions \\
\hline
\hline
 & $Z_{e,g^k} = 
|\chi^{}_0|^2 + \zeta^{2k} \, |\chi^{}_1|^2 + \zeta^{4k} \, |\chi^{}_2|^2 + \zeta^{6k}
\, |\chi^{}_3|^2 + \zeta^{8k} \, |\chi^{}_4|^2 + \zeta^{10k} \, |\chi^{}_5|^2$ \\ 
$E^{}_8$ & $Z_{g,g^k} = \zeta^{7k} \, \chi^*_0\chi^{}_1 + \zeta^{9k} \, \chi_1^*\chi^{}_2 +
\zeta^{11k} \, \chi_2^*\chi^{}_3 + \zeta^{k} \, \chi_3^*\chi^{}_4 + 
\zeta^{3k} \, \chi_4^*\chi^{}_5 + \zeta^{5k} \, \chi_5^*\chi^{}_0$ \\
& $Z_{g^2,g^k} = \zeta^{2k} \, \chi_0^*\chi^{}_2 + \zeta^{4k} \, \chi_1^*\chi^{}_3 +
\zeta^{6k} \, \chi_2^*\chi^{}_4 + \zeta^{8k} \, \chi_3^*\chi^{}_5 + 
\zeta^{10k} \, \chi_4^*\chi^{}_0 + \chi_5^*\chi^{}_1$ \\
& $Z_{g^3,g^k} = \zeta^{9k} \, \chi_0^*\chi^{}_3 + \zeta^{11k} \, \chi_1^*\chi^{}_4 +
\zeta^{k} \, \chi_2^*\chi^{}_5 + \zeta^{3k} \, \chi_3^*\chi^{}_0 + 
\zeta^{5k} \, \chi_4^*\chi^{}_1 + \zeta^{7k} \, \chi_5^*\chi^{}_2$ \\
& $Z_{g^4,g^k} = \zeta^{4k} \, \chi_0^*\chi^{}_4 + \zeta^{6k} \, \chi_1^*\chi^{}_5 +
\zeta^{8k} \, \chi_2^*\chi^{}_0 + \zeta^{10k} \, \chi_3^*\chi^{}_1 + 
 \chi_4^*\chi^{}_2 + \zeta^{2k} \, \chi_5^*\chi^{}_3$ \\
& $Z_{g^5,g^k} = \zeta^{11k} \, \chi_0^*\chi^{}_5 + \zeta^{k} \, \chi_1^*\chi^{}_0 +
\zeta^{3k} \, \chi_2^*\chi^{}_1 + \zeta^{5k} \, \chi_3^*\chi^{}_2 + 
\zeta^{7k} \, \chi_4^*\chi^{}_3 + \zeta^{9k} \, \chi_5^*\chi^{}_4$ \\
$\matrix{ & \cr & \cr & \cr}$ & with 
\begin{tabular}[t]{ll}
$\chi_0 = \chi^{}_{(1,1)}+\chi^{}_{(3,3)}$ &
\qquad $\chi_1 = \chi^{}_{(1,3)}+\chi^{}_{(4,3)}$ \\
$\chi_2 = \chi^{}_{(2,3)}+\chi^{}_{(6,1)}$ &
\qquad $\chi_3 = \chi^{}_{(1,4)}+\chi^{}_{(4,1)}$ \\
$\chi_4 = \chi^{}_{(1,6)}+\chi^{}_{(3,2)}$ &
\qquad$\chi_5 = \chi^{}_{(3,1)}+\chi^{}_{(3,4)}$\\
\end{tabular}\\
\hline
$E^*_8$ & All partition functions are obtained from the above via $C$\\
\hline
\end{tabular}
\end{center}
\caption{\footnotesize Consistent set of partition functions for the $E^{}_8$
and $E^{*}_8$ models, frustrated by a $\Z_6$ group of symmetry, realized
projectively. The number $\zeta$ is a primitive 12--th root of unity.}
\end{table}


If one performs modular transformations on it, one finds that the action of $g$ is by
sixth roots of unity in the sectors ${\cal H}_{e,g^2,g^4}$, and by twelveth roots of 1
in ${\cal H}_{g,g^3,g^5}$, more precisely by $i$ times sixth roots of 1, a clear sign
that projective representations are present in those three sectors. The projective or
non--projective nature of the representations in the various sectors must however obey
the consistency conditions set by the modular transformations. One should be able to
find six cocycles $\omega_g$, one for each sector which determines the nature of the
$\Z_6$ representations in that sector, such that the transformation laws (\ref{pmod})
are fulfilled
\beq
Z_{g,g'}(\tau) = \omega_g (g,g') \; Z_{g,gg'}(\tau + 1) = 
\omega_{g'}^{-1}(g,g^{-1}) \; Z_{g',g^{-1}} ({\textstyle -{1 \over \tau}}).
\label{ppmod}
\eeq

We note that in general the determination of the cocycles affects that of the partition
functions and vice-versa: the functions $Z_{g,g'}$ contain the full information
about the cocycles $\omega_g$, but cannot be computed unless some cocycles are given.
As a consequence, more than one consistent set of partition functions and of cocycles
can be found (except for a $\Z_2$ group). One may also observe that a number of
partition functions can be determined from $Z_{e,g^k}$ without the knowledge of any
cocycle, namely all $Z_{g^k,e}$ and the diagonal ones $Z_{g^k,g^k}$. Moreover, a
limited number of cocycle values are needed to compute the full table of partition
functions. 

The simplest solution to (\ref{ppmod}) is as follows. As the group acts in the 
sectors ${\cal H}_{e,g^2,g^4}$ by sixth roots of unity, there is no need to introduce a
non--trivial cocycle there. So we put $\omega_e = \omega_{g^2} = \omega_{g^4} = 1$.
On the other hand, the function $Z_{g,g}$ says that $g$ acts in ${\cal H}_g$ by $i$
times sixth roots of unity. The simplest to assume is that $g^k$ acts as the
$k$--th power of $g$, and this fixes the cocycle $\omega_g$ to be
$\omega_g(g^k,g^l) = i^{k+l-\langle k+l \rangle_6}$, where $\langle n \rangle_6$ stands
for the residue of $n$ modulo 6, taken between 0 and 5. In turn, this allows to compute
all $Z_{g,g^k}$ and then all $Z_{g^k,g}$. The same assumption for ${\cal H}_{g^3,g^5}$
as for ${\cal H}_{g}$ (the action of $g^k$ is the $k$--th power of the action of $g$)
yields the same cocycle, so that all together 
\beq
\begin{array}{l}
\omega_e(g^k,g^l) = \omega_{g^2}(g^k,g^l) = \omega_{g^4}(g^k,g^l) = +1, \\
\omega_g(g^k,g^l) = \omega_{g^3}(g^k,g^l) = \omega_{g^5}(g^k,g^l) = i^{k+l-\langle
k+l \rangle_6}.
\end{array}
\eeq

They determine non--ambiguously all partition functions, given in Table 5, which
display a neat cyclic structure. Finally, one has to make sure that our assumptions
are self--consistent by verifying that the transformations (\ref{ppmod}) are satisfied
for all $g,g'$, which they are.  

The same analysis can be made for the $E^*_8$ model, starting from the graph ${\cal
E}^*_8$. It has a $\Z_2$ automorphism which leads to a projective $\Z_2$ symmetry in
the corresponding field theory, and eventually to a projective $\Z_6$.

\subsubsection{The $E^{}_{24}$ invariant of su(3)}

This is the third and last invariant of $su(3)$ which is compatible with a projective
symmetry. A numerical analysis, using the Galois symmetry, shows that the field theory
is not compatible with a symmetry group acting by true representations. However, guided
by the corresponding graph ${\cal E}_{24}$, reproduced below, one can see that it is
compatible with a $\Z_2$ projective action. 

\begin{figure}[htb]
\leavevmode
\begin{center}
\mbox{\epsfscale 500 \epsfbox{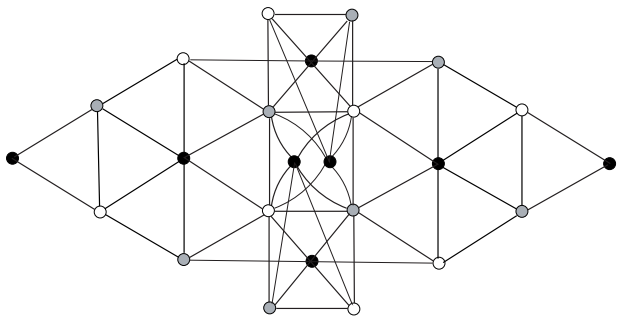}}
\end{center}
\caption{\footnotesize The graph ${\cal E}_{24}^{}$ corresponding to the level 21
$su(3)$ modular invariant $E_{24}^{}$.}  
\end{figure}

The only non--trivial automorphism of the graph is an inversion through its center,
and preserves the colour of the nodes. Acting on the eigenvectors of the adjacency
matrix, it has twelve eigenvalues equal to $+1$ and twelve eigenvalues equal to $-1$.
All eigenvalues $+1$ correspond to the twelve characters contained in the block of the
identity, from which one sets
\beq
Z_{e,g} = |\chi^{}_0|^2 - |\chi^{}_1|^2,
\eeq
with
\bea
&& \chi^{}_0 = \chi^{}_{(1,1)} + \chi^{}_{(5,5)} + \chi^{}_{(11,11)} + \chi^{}_{(7,7)} +
\hbox{$\mu$ and $\mu^2$ rotations}, \nonumber \\
&& \chi^{}_1 = \chi^{}_{(1,7)} + \chi^{}_{(7,1)} + \chi^{}_{(5,8)} + \chi^{}_{(8,5)} + 
\hbox{$\mu$ and $\mu^2$rotations}.
\eea

One then computes, via an $S$ and a $T^{-1}$ modular transformation, that
\beq
Z_{g,e} = \chi^*_0 \, \chi^{}_1 + \chi^*_1 \, \chi^{}_0\,, \qquad
Z_{g,g} = -i \, \chi^*_0 \, \chi^{}_1 + i \, \chi^*_1 \, \chi^{}_0\,.
\eeq

One may note that, as follows from (\ref{ppmod}), no cocycle is required to compute
these partition functions. A posteriori, one checks that the relations (\ref{ppmod}) are
indeed verified provided $\omega_e=1$, $\omega_g(g,g)=-1$ and all other values equal to
1. This cocycle is universal for projective $\Z_2$ actions (in the $A_{n-1}$ models of
$su(2)$ for $n$ odd, see Table 1, and in the $E_8^{(*)}$ models of $su(3)$).


\section{Proofs}

We give in this section the elements needed to prove the results announced in
Tables 1 to 3, concerning the maximal symmetry of each model and the corresponding
partition functions.

Modular transformations form clearly the most important ingredient. Affine
characters $\chi_p(\tau)$, labelled by some finite set $P_{++}$,
transform linearly under the modular group. Under the two fundamental transformations,
generating the whole group, the characters transform as \cite{kac}
\beq
\chi_p({\textstyle {-1 \over \tau}}) = \sum_{p' \in P_{++}} \;
S_{ p, p'} \, \chi_{p'}(\tau), \qquad
\chi_{ p}(\tau + 1) = \sum_{ p' \in P_{++}} \; T_{ p, p'} \,\chi_{ p'}(\tau).
\eeq
The two matrices $S$ and $T$, symmetric and unitary, are the essential tools to compute
the modular transformations of the partition functions. They generate a representation
of (in general) the double covering ${\rm SL}_2(\Z$) of the modular group. Concrete
expressions for $S$ and $T$ are given in \cite{kac} for all affine Lie algebras, and
will be reproduced below in the case of $\widehat{su}(2)$ and $\widehat{su}(3)$. These
two matrices, especially $S$, possess fascinating and useful symmetry properties under
the action of the universal Galois group Gal($\bar\Q/\Q)$ \cite{cg}. We will use them
on several occasions, mainly in isolated computer--assisted cases.

\subsection{The su(2) theories}

The integrable representations of the affine Lie algebra $\widehat{su}(2)_k$, with
$k$ the level, a non--negative integer, can be labelled by $su(2)$ highest weights
$p$ in $P_{++}^{(n)} = \{p \in \Z \;:\; 1\leq p \leq n-1\}$ where the level has been
traded for the height $n = k+2$. The matrices $S$ and $T$ are given explicitely by
\beq
S_{p,p'} = \sqrt{2 \over n} \, \sin{\pi pp' \over n}, \qquad
T_{p,p'} = {\rm e}^{2i\pi ({p^2 \over 4n}-{1 \over 8})} \, \delta_{p,p'}.
\eeq

By definition, simple currents correspond to those weights $J$ such that $S_{1,J} =
S_{1,1}$. In affine $su(2)$ theories, there are two of them, given by $J_0=1$ and
$J_1=n-1$. The simple current $J_1$ generates the second order automorphism $\mu$ of
$P^{(n)}_{++}$, given $\mu(p) = n-p$, with respect to which the matrix $S$ transforms as
\beq
S_{\mu^k(p),\mu^\ell(p')} = (-1)^{k(p'+1) + \ell(p+1) + k\ell n} \, S_{p,p'},  \qquad
a,b=0,1.
\label{sc2}
\eeq

The proof of the results in Table 1 follows closely the one given in \cite{rv} for the
same problem in the minimal models (and is even simpler). In particular, the arguments
there show that the $su(2)$ theories have a maximal symmetry at most equal to $\Z_2$,
except the $D_{4m+2}$ models whose maximal symmetry is a subgroup of $\Z_{30}$, this
last result making use of the Galois symmetry of $S$. The rest of the proof can be
easily adapted from \cite{rv}. As illustration, we give the detailed proof for the
$D_{4m+4}$ theories. The same method can be used for the corresponding series
$D^{(*)}_n$, $n \neq 0 \bmod 3$, of $su(3)$, and for the diagonal theories, or indeed for any simple current automorphism modular invariant. 

The starting point is the modular invariant $D_{{n \over 2}+1}$, with $n=0 \bmod 4$,
\beq
Z_{e,e} = \sum_p \; \chi^*_p \, \chi^{}_{\mu^{p+1}(p)} = 
\sum_{p=1,\, {\rm odd}}^{n-1}\, |\chi_{p}|^2 +
\sum_{p=2,\, {\rm even}}^{n-2}\, \chi_{p}^*\,\chi^{}_{n-p},
\label{dser2}
\eeq
which we suppose is compatible with a $\Z_N$ symmetry.

The matrix $M_{p,p'}^{(e)}(e)$ is a permutation matrix and thus $M_{p,p'}^{(e)}(g)$,
specifying the action of a $\Z_N$ generator in the periodic sector, is a phased
permutation, with entries equal to $N$--th roots of 1. Whatever these phases are, its
$2N$--th power is equal to the identity. The modular transformation
\beq
M^{(g)}(e) = S^\dag \, M^{(e)}(g) \, S,
\eeq
shows that the same is true of $M^{(g)}(e)$ which, being positive integer--valued,
must be a permutation matrix. Let $M_{p,p'}^{(g)}(e) = \delta_{p,\pi(p')}$.

The entry $p=p'=1$ of the previous equation, written as $S M^{(g)}(e) =
M^{(e)}(g) S$, shows that $\pi(1) = n-1$ must be the non--trivial simple current
($\pi(1)$ cannot be 1, because the twisted sector does not contain the identity
field). Taking the same equation again for arbitrary $p$ and $p'=1$, one finds, from
(\ref{sc2}), that the phases are in fact equal to signs, implying $N=2$. More
precisely, one finds the explicit form $M_{p,p'}^{(e)}(g) = (-1)^{p+1}
\, \delta_{p',\mu^{p+1}(p)}$. This in turn determines $M^{(g)}(e)$ as well as
$M_{p,p'}^{(g)}(g)$, as given in Table 1. 

This shows that $\Z_2$ is the only cyclic symmetry compatible with the modular
invariant (\ref{dser2}), and that it has a unique realization in the periodic sector,
completing the proof that it is the maximal symmetry. \cqfd


\subsection{The su(3) theories}

The characters of $\widehat{su}(3)_k$ are indexed by $su(3)$ dominant weights in
$P_{++}^{(n)} = \{p=(a,b) \in \Z^2 \;:\; 1 \leq a,b,a+b \leq n-1\}$, with the height
defined by $n=k+3$. The modular matrices read 
\bea
&& S_{p,p'} = \frac{-i}{\sqrt{3}\, n}\;\sum_{w \in W(su(3))}\,(\det w)\,e^{-2i\pi
(w(p)\cdot p')/n}, \\
\noalign{\smallskip}
&& T_{p,p'} = e^{2i\pi(a^2+b^2+ab-n)/3n}\,\delta_{p,p'},
\eea
where the $w$ summation is over the Weyl group of $su(3)$. They satisfy $S^2 = (ST)^3 =
C$, where the CFT charge conjugation coincides with $su(3)$ conjugation,
$C(a,b)=(b,a)$. In particular $S^*_{p,p'} = S^{}_{C(p),p'} = S^{}_{p,C(p')}$.

When one of the indices of $S$ is a diagonal weight, the expression simplifies to
\beq
S_{(l,l),(a,b)} = \frac{8}{n\,\sqrt{3}}\,\sin[{\pi al\over n}] \, \sin[{\pi bl\over
n}] \,\sin[{\pi (a+b)l\over n}].
\eeq

There are three simple currents $J$, satisfying as before $S_{(1,1),J} =
S_{(1,1),(1,1)}$, given by $J_0=(1,1)$, $J_1=(n-2,1)$ and $J_2=(1,n-2)$. The last two
generate order 3 automorphisms of $P_{++}^{(n)}$, given by $\mu(a,b) = (n-a-b,a)$ and
$\mu^2(a,b) = (b,n-a-b)$, under which $S$ transforms as
\beq
S_{\mu^k(p),\mu^\ell(p')} = {\rm e}^{2i\pi(kt(p') + \ell t(p) + k\ell n)/3} \,
S_{p,p'},  \qquad k,\ell=0,1,2,
\label{smu}
\eeq
with $t(a,b)=a-b \bmod 3$ the triality.

The automorphism modular invariants of $su(3)$ can be handled by the method detailed in
the previous section in the case of $su(2)$, while the few exceptional invariants can
be analyzed on a case--by--case basis, with the results given in Table 2. The remaining
series $D_{n}^{(*)}$, with $n=0 \bmod 3$, is more peculiar and must be treated
separately. It could in principle be handled by the same methods as in \cite{rv}, that
relied on the explicit solution of the constraints imposed by the Galois symmetries of
$S$. However, in $su(3)$, this method is tedious, and is in any way useless
in other cases. Therefore we have chosen a proof that is independent of Galois
arguments. 

The rest of this section is devoted to the proof that the $D_{n}^{(*)}$ modular
invariants, $n=0 \bmod 3$, are compatible with the cyclic symmetry $\Z_3$, and only
that one if $n \geq 9$, realized in a unique way if $n \geq 12$, and in two different
ways if $n=6,9$. 

\bigskip
The modular invariant partition function reads
\beq
Z_{0,0} = \sum_{p\,:\,t(p)=0}  \bigg[\sum_{j=0}^{2} \;
\chi^*_{p}\,\chi^{}_{\mu^{j}(p)} \bigg] = {\textstyle 1 \over 3} \sum_{p\,:\,t(p)=0}
|\chi_p + \chi_{\mu(p)} + \chi_{\mu^2(p)}|^2.
\eeq
All fields appear with a multiplicity equal to 1, except $[({n \over 3},{n \over 3}), 
({n \over 3},{n \over 3})]$ which occurs with multiplicity 3.  

So we look for cyclic symmetries and assume the compatibility of the above modular
invariant with a $\Z_N$ symmetry. We want to show first that $N$ must be equal to 3 if
$n \geq 9$. In order to simplify the notations, we write $M^{(i,j)}$ instead of
$M^{(g^i)}(g^j)$ for $g$ a generator of $\Z_N$. We mainly concentrate on
$M^{(1,0)}$ and $M^{(0,1)}$.

A preliminary but very useful observation is that $M_{\mu^{k}(p) , \mu^{\ell}(p')}
^{(1,0)} = M_{p,p'}^{(1,0)}$, as a simple consequence of (\ref{smu}) and the fact that
$M^{(0,1)}_{p,p'}$ is zero for all $p,p'$ of non--zero triality.

The modular relation $SM^{(1,0)} = M^{(0,1)}S$ implies
\beq
\sum_p \; S^{}_{(1,1),p} \, M_{p,(1,1)}^{(1,0)} = \sum_p \; M_{(1,1),p}^{(0,1)} 
\, S^{}_{p,(1,1)} = (1+2 \cos{2\pi q \over N}) \, S^{}_{(1,1),(1,1)} \leq 3 
S_{(1,1),(1,1)}.
\label{ineq}
\eeq
Using again the symmetry (\ref{smu}) and the condition $M_{(1,1),(1,1)}^{(1,0)}=0$, and
remembering that $S_{(1,1),p} \ge S_{(1,1) (1,1)}$, we see that the only way the above
inequality can be satisfied is that the whole column $M^{(1,0)}_{p,(1,1)}$ is equal
to zero, except possibly for the entry corresponding to the fixed point $M^{(1,0)}_{({n
\over 3},{n \over 3}),(1,1)} \leq 2$.

A non--zero value $M^{(1,0)}_{({n \over 3},{n \over 3}),(1,1)} = m > 0$ has however to
satisfy the above inequality, which explicitely requires
\beq
m \, \Big(\sin{\frac{\pi}{3}}\Big)^2 \,\sin{\frac{2 \pi}{3}} \leq
3 \, \Big(\sin{\frac{\pi}{n}}\Big)^2 \,\sin{\frac{2 \pi}{n}},
\eeq
which holds for $n=6$ and $m=1$ only. Feeding this back in (\ref{ineq}) then yields
$q=0$ and no restriction on $N$ at this stage. Thus the isolated case $n=6$ demands a
separate treatment which we will not detail here, the results being summarized in Table
3. 

Thus we may assume $M_{p,(1,1)}^{(1,0)} = 0$ for all $p$ (the arguments that follow are
valid for all $n \geq 6$). Then Eq. (\ref{ineq}) forces $\cos\frac{2 \pi q}{N} = -
\frac{1}{2}$, which  implies $N$ divisible by 3, and hence, for all $p$ of zero
triality,
\beq
\sum_{p'} \, S^{}_{p,p'} \, M_{p',(1,1)}^{(1,0)} = 0 =  \sum_{p'} \, M_{p,p'}^{(0,1)}
\;  S^{}_{p',(1,1)}.
\eeq
Since $S_{p',(1,1)} \neq 0$, one obtains $\sum_k \, M_{p,\mu^k(p)}^{(0,1)} =
0$, implying in particular $M^{(0,1)}_{({n \over 3},{n \over 3}), ({n \over 3},{n \over
3})} = 0$.

The same arguments with the equation $M^{(1,0)} S^\dagger = S^\dagger M^{(0,1)}$ give
similar constraints for the columns of $M^{(0,1)}$, namely that $\sum_k \,
M_{\mu^k(p),p}^{(0,1)} = 0$.

To prove that $N=3$, one can repeat the same calculations for any matrix $M^{(0,x)}$,
with $x$ integer between 1 and $N-1$. Because $g$ acts on the non--degenerate fields by
multiplication by a phase $\zeta$, so does $g^x$, by a phase equal to $\zeta^x$.
Therefore $M^{(0,x)}_{p,p'} = [M^{(0,1)}_{p,p'}]^x$ for all pairs $(p,\mu^j(p))$ and $p
\neq ({n \over 3},{n \over 3})$. It is now straightforward to see that the constraints
which will follow from this are
\beq
\sum_{k=0}^2 \, [M_{p,\mu^k(p)}^{(0,1)}]^x = \sum_{k=0}^2 \, [M_{\mu^k(p),p}^{(0,1)}]^x
= 0, \qquad \hbox{for all} \;  1 \leq x \leq N-1.
\label{sumsum}
\eeq

The previous equations for $x=1$ show that the three numbers $M_{p,p}^{(0,1)}$,
$M_{p,\mu(p)}^{(0,1)}$ and $M_{p,\mu^2(p)}^{(0,1)}$ are the three distinct third roots
of unity, up to a global $N$--th root of 1. 

Now if $N \geq 6$, one could take $x = 3$, for which (\ref{sumsum}) is clearly
violated. Thus we find that \emph{the only cyclic symmetry compatible with the
modular invariants $D_{n}$, $n \geq 9$ divisible by 3, is a $\Z_{3}$ symmetry}.

\bigskip
The next step is to determine the possible realizations of this $\Z_{3}$ group, and in
fact to show that there is a unique realization when $n \geq 12$, and two realizations 
for $n=6,9$, related to each other by conjugation $C$.

From the constraints derived in the previous step, one knows that $M^{(0,1)}_{({n
\over 3},{n \over 3}), ({n \over 3},{n \over 3})} = 0$ and that the 3--by--3 block of
$M^{(0,1)}$ containing the identity field has row and column sums equal to zero,
and therefore has one of the four forms
\beq
\mbox{\footnotesize{(a)=$\pmatrix{1& \omega  & \omega^2\cr
\omega^2 & 1 & \omega \cr
\omega & \omega^2 & 1 }$, \hspace{3mm}
(b)=$\pmatrix{1& \omega^2 & \omega \cr
\omega^2 & \omega & 1 \cr
\omega & 1& \omega^2 \cr}$, \hspace{3mm}
(c)=$\pmatrix{1& \omega^2 & \omega \cr
\omega & 1 & \omega^2 \cr
\omega^2 & \omega & 1}$, \hspace{3mm}
(d)=$\pmatrix{1& \omega & \omega^2 \cr
\omega & \omega^2 & 1 \cr
\omega^2 & 1 & \omega }$}},
\label{blocks}
\eeq
where the rows and columns are labelled by $(1,1)$, $\mu(1,1)$ and $\mu^2(1,1)$, and
where $\omega = e^{2 i \pi /3}$.

We note that the blocks (c) and (d) are related to (a) and (b) respectively by the
change $\omega \leftrightarrow \omega^2$, equivalent to a change of generator of
$\Z_3$. We can therefore omit them. We will show that the form (a) for the block of the
identity uniquely determines all $\Z_3$ partition functions, and that the form (b)
leads to a contradiction unless $n=6,9$.

\medskip
Assume that the block of $M^{(0,1)}$ of the identity is (a). Then for any $p \in P_{++}$
\beq
\sum_{p'} \, S^{}_{(1,1),p'} \, M^{(1,0)}_{p',p} = \sum_{p'} \, M^{(0,1)}_{(1,1),p'} \,
S^{}_{p',p} = [1 + \omega^{1+t(p)} + \omega^{2(1+t(p))}] \, S_{(1,1),p}.
\label{416}
\eeq
The r.h.s. is equal to zero if $t(p)=0$ or 1, in which case the l.h.s. implies
$M^{(1,0)}_{p',p} = 0$ for all weights $p$ such that $t(p)=0,1$. 

Likewise, the relations
\beq
\sum_{p'} \, M^{(1,0)}_{p,p'} \, S^\dagger_{p',(1,1)} = \sum_{p'} \,
S^\dagger_{p,p'} \, M^{(0,1)}_{p',(1,1)} = [1 + \omega^{2-t(p)} + \omega^{2(2-t(p))}] 
\, S^\dagger_{p,(1,1)}
\label{417}
\eeq
show $M^{(1,0)}_{p,p'} = 0$ for all weights $p$ such that $t(p)=0,1$. 

So altogether, one finds that $M^{(1,0)}$ is non--zero on the triality 2 weights only.
For those, the previous two equations yield
\beq
\sum_{p'} \, S^{}_{(1,1),p'} \, M^{(1,0)}_{p',p} = \sum_{p'} \, M^{(1,0)}_{p,p'} \,
S^{}_{p',(1,1)} = 3 S_{(1,1),p}\, , \qquad \qquad t(p)=2.
\label{t2}
\eeq
They are clearly satisfied if we set $M^{(1,0)}_{p,p'} = 1$ if $p'=p,\mu(p)$ or
$\mu^2(p)$, and 0 otherwise. We show that this is in fact the only solution. 

It is certainly true for the second fundamental weight $\lambda^2=(1,2)$ and its two
partners $\mu(\lambda^2),\, \mu^2(\lambda^2)$, because, in the subset of weights with
triality 2, they have the smallest value of $S_{(1,1),p}$ (and so of quantum dimension)
\cite{grw}. Therefore, $M^{(1,0)}_{\mu^k(\lambda^2),p} =
M^{(1,0)}_{p, \mu^\ell(\lambda^2)} = 1$ if $p \in \{\lambda^2, \mu(\lambda^2),
\mu^2(\lambda^2)\}$, and 0 if $p$ is anything else.

Writing once more the equation $SM^{(1,0)} = M^{(0,1)}S$ for an arbitrary $p$ of zero
triality and $\lambda^2$, one obtains
\beq
\sum_{p'} \, S_{p,p'} \, M^{(1,0)}_{p',\lambda^2} = 3S_{p,\lambda^2} = \sum_{k=0}^2 \,
M^{(0,1)}_{p,\mu^k(p)} \, S_{\mu^k(p),\lambda^2} = \sum_{k=0}^2 \, \omega^{2k}\,
M^{(0,1)}_{p,\mu^k(p)} \, S_{p,\lambda^2}.
\eeq
As $S_{p,\lambda^2} \neq 0$ for all $p \neq ({n \over 3},{n \over 3})$, one deduces that
$M^{(0,1)}_{p,\mu^k(p)} = \omega^k$ for all such $p$, or in other words, that all
3--by--3 blocks of $M^{(0,1)}$ are the same and equal to the matrix (a). Since we
already know that $M^{(0,1)}_{({n \over 3},{n \over 3})} = 0$, one has
\beq
Z_{e,g} = {\textstyle 1 \over 3} \sum_{p\,:\,t(p)=0} |\chi_p + \omega \chi_{\mu(p)} +
\omega^2 \chi_{\mu^2(p)}|^2.
\eeq
Its various modular transformations fill the table of partition functions, given in
Table 2. Thus, when the block of $M^{(0,1)}$ containing the identity is given by the
matrix (a) in (\ref{blocks}), there is a unique realization of the $\Z_3$ symmetry,
for all $n \geq 6$ (divisible by 3). 

\medskip
There is a second independent possibility for that block, namely (b). Since the first
columns of (a) and (b) are equal, Eq. (\ref{417}) remains, while Eq. (\ref{416}) only
slightly changes, to the effect that now only the columns of $M^{(1,0)}$ labelled by
weights of triality 1 are non--zero, whereas the rows labelled by weights of triality
2 are non--zero. Equivalently, defining $\widetilde M^{(1,0)}_{p,p'} =
M^{(1,0)}_{p,C(p')}$, one finds that $\widetilde M^{(1,0)}$ satisfies all the conditions
that $M^{(1,0)}$ satisfied in case (a), namely
\beq
\sum_{p'} \, S^{}_{(1,1),p'} \, \widetilde M^{(1,0)}_{p',p} = \sum_{p'} \,
\widetilde M^{(1,0)}_{p,p'} \, S^{}_{p',(1,1)} = 3 S_{(1,1),p} \, \delta_{t(p),2}.
\eeq

The same reasoning as above determines the same unique $\widetilde M^{(1,0)}$, from
which we deduce that the matrix $M^{(1,0)}$ in case (b) is the $C$--conjugate of that in
case (a): $M^{(1,0),(b)} = M^{(1,0),(a)}\,C$. Then an inverse $S$ modular
transformation gives us at once that $M^{(0,1),(b)} = M^{(0,1),(a)}\,C$, and a
partition function $Z_{e,g}$ which is the $C$--conjugate of that of case (a). This form
for $Z_{e,g}$ is not consistent with the modular invariant $D_n$ we started
from, unless that invariant is self--conjugate, that is, for $n=6,9$. 

This concludes our proof for the invariants $D_n$, $n=0 \bmod 3$. For $n \geq 12$, there
is only one realization of a symmetry $\Z_3$, while there are two for $n=6,9$, conjugate
of each other. Furthermore, there is some room for other cyclic symmetries when $n=6$,
and a separate analysis of this particular case furnishes the results of Section 3.3.1.
\cqfd


\section{Conclusion}

The first purpose of this article was to determine the (finite) symmetries of affine
conformal theories based on $su(2)$ and $su(3)$, by using the modular covariance of the
torus partition functions. The results, in the form of a list of groups and of partition
functions which specify the contents of the twisted sectors and the way these groups
act on the fields, have been reported in Section 3.1, see also the Tables 1 to 3. 

However a strong motivation for this work was to see if the symmetries present in the
field theories, and the representations carried by the various sectors, are in some
way encoded in the graphs that have been associated with these theories. Indeed the many
points of view that have been taken over the last ten years have consistently shown
that these graphs govern many fundamental aspects of those models. Hence our second
purpose was to examine and to probe the relevance of the graphs from symmetry
considerations. 

In this respect, a rather firm conclusion is that, as expected, the graphs indeed have
much to say about the symmetries and their realizations in the field theories. This is
especially true for the affine $su(2)$ models, where the matching is complete
(provided one allows for projective representations). Surprisingly perhaps, this is
less so for the $su(3)$ models, where in many cases the graphs have symmetries
(automorphisms) unmatched in the field theories. However, taken in the other way, the
connection works nicely and universally, since a symmetry in the field theory always
has a counterpart in the graph (except for the non--colourable graphs in
$su(3)$, for known reasons). In addition the content of the twisted sectors and the way
the symmetry is represented in them can be recovered from the graph, thereby extending
what the graphs had been devised for in the first place, namely the coding of the
diagonal terms in a modular invariant. Most of these features have remained mere
observations. 

The investigation of the symmetry features of these models in a cylindric geometry 
would certainly form a natural continuation of this work. 


\vskip 1truecm
\section*{Acknowledgments}

P.R. thanks Jean-Bernard Zuber for useful discussions about the su(3) graphs.
We also thank the authors of \cite{bppz} for the permission to use their eps codes 
for the su(3) graph figures. 



\begin{thebibliography}{99}

\bibitem{bpz}
A.A. Belavin, A.M. Polyakov and A.B. Zamolodchikov, Nucl. Phys. B 241 (1984) 333.

\bibitem{g}
K. Gaw\c edzki, {\it Conformal field theory: a case study}, {\tt hep-th/9904145}.

\bibitem{w}
M. Walton, {\it Affine Kac--Moody Algebras and the Wess--Zumino--Witten Model}, {\tt
hep-th/9911187}.

\bibitem{ciz} A. Cappelli, C. Itzykson and J.-B. Zuber, Commun. Math. Phys. 113 (1987)
1.

\bibitem{kato} A. Kato, Mod. Phys. Lett. A 2 (1987) 585.

\bibitem{gannon}
T. Gannon, Annales Inst. Poincar\'e: Phys. Th\'eor. 65 (1996) 15.

\bibitem {zuber}
J.-B. Zuber, Phys. Lett. B 176 (1986) 127.

\bibitem {rv}
P. Ruelle, O. Verhoeven, Nucl. Phys. B 535 (1998) 650.

\bibitem{r}
P. Ruelle, J. Phys. A: Math. Gen. 32 (1999) 8831.

\bibitem{kac} V. Kac, {\it Infinite Dimensional Lie algebras}, (Cambridge University
Press, Cambridge, 1990).

\bibitem{cardy2}
J. Cardy, Nucl. Phys. B 275 [FS17] (1986) 200.

\bibitem{iz}
C. Itzykson and J.-B. Zuber, Nucl. Phys. B 275 [FS17] (1986) 580.

\bibitem{bppz}
R.E. Behrend, P.A. Pearce, V.B. Petkova and J.-B. Zuber, {\it Boundary
Conditions in Rational Conformal Field Theories}, {\tt hep-th/9908036}.

\bibitem{dms}
P. Di Francesco, P. Mathieu and D. S\'en\'echal, {\it Conformal Field Theory},
(Springer, New York, 1997).

\bibitem{ginsp}
P. Ginsparg, {\it Applied conformal field theory}, in {\it Les Houches, session XLIX,
``Fields, Strings and Critical Phenomena''}, eds E. Br\'ezin and J. Zinn-Justin 
(Elsevier, New York, 1989).

\bibitem{vafa}
C. Vafa, Nucl. Phys. B 273 (1986) 592.

\bibitem{dfz}
P. Di Francesco and J.-B. Zuber, Nucl. Phys. B 338 (1990) 602.

\bibitem{df}
P. Di Francesco, Int. J. Mod. Phys. A 7 (1992) 407.

\bibitem{fs2}
J. Fuchs and C. Schweigert, Nucl. Phys. B 530 (1998) 99.

\bibitem{z2}
J.-B. Zuber, {\it CFT, BCFT, ADE and all that}, {\tt hep-th/0006151}.

\bibitem{abf}
G.E. Andrews, R.J. Baxter and J.P. Forrester, J. Stat. Phys. 35 (1984) 193.

\bibitem{pas}
V. Pasquier, Nucl. Phys. B 285 (1987) 162.

\bibitem{wns}
S. Warnaar, B. Nienhuis and K. Seaton, Phys. Rev. Lett. 69 (1992) 710.

\bibitem{o'bp}
D. O'Brien and P.A. Pearce, J. Phys. A: Math. Gen. 28 (1995) 4891.

\bibitem{bp}
R.E. Behrend and P.A. Pearce, {\it Integrable and Conformal Boundary Conditions for
$\widehat{sl}(2)$ $A$-$D$-$E$ Lattice Models and Unitary Minimal Conformal Field
Theories}, {\tt hep-th/0006094}.

\bibitem{pz}
V.B. Petkova and J.-B. Zuber, Nucl. Phys. B 438 (1995) 347; Nucl. Phys. B 463 (1996)
161.

\bibitem{bek}
J. B\"ockenhauer, D.E. Evans and Y. Kawahigashi, Commun. Math. Phys. 208 (1999), 429;
Commun. Math. Phys. 210 (2000) 733. 

\bibitem{cardy3}
J. Cardy, Nucl. Phys. B 324 (1989) 581.

\bibitem{pss}
G. Pradisi, A. Sagnotti and Ya.S. Stanev, Phys. Lett. B 381 (1996) 97.

\bibitem{fs1}
J. Fuchs and C. Schweigert, Phys. Lett. B 414 (1997) 251.

\bibitem{k}
I. Kostov, Nucl. Phys. B 300 (1988) 559.

\bibitem{tw}
Thomas and Wood, {\it Group Tables}, (Shiva Publishing Limited, 1980).

\bibitem{karp}
G. Karpilovsky, {\it Projective representations of finite groups}, (Marcel Dekker, New
York, 1985). 

\bibitem{xu}
F. Xu, Commun. Math. Phys. 192 (1998) 349.

\bibitem{cg}
A. Coste and T. Gannon, Phys. Lett. B 323 (1994) 316.

\bibitem{grw}
T. Gannon, P. Ruelle and M. Walton, Commun. Math. Phys. 179 (1996) 121.


\end{thebibliography}
\end{document}